\documentclass[fleqn,usenatbib]{mnras}

\usepackage{newtxtext,newtxmath}
\usepackage{supertabular}

\usepackage[T1]{fontenc}

\DeclareRobustCommand{\VAN}[3]{#2}
\let\VANthebibliography\thebibliography
\def\thebibliography{\DeclareRobustCommand{\VAN}[3]{##3}\VANthebibliography}

\usepackage{graphicx}	
\usepackage{amsmath}	

\usepackage{rotating}
\usepackage{bm}
\usepackage{longtable,tabu}
\usepackage{xcolor}
\usepackage{pifont}
\usepackage{soul}
\usepackage{ulem}

\usepackage{graphicx,pifont}
\let\oldding\ding
\renewcommand{\ding}[2][1]{\scalebox{#1}{\oldding{#2}}}

\usepackage{array}
\newcolumntype{H}{>{\setbox0=\hbox\bgroup}c<{\egroup}@{}}

\title[Cepheids in open clusters using Gaia data]{A revisited study of Cepheids in open clusters in the Gaia era}

\author[Medina, Lemasle \& Grebel]{
Gustavo E. Medina,$^{1}$\thanks{E-mail: gustavo.medina@uni-heidelberg.de}
Bertrand Lemasle,$^{1}$
Eva K. Grebel$^{1}$
\\
$^{1}$Astronomisches Rechen-Institut, Zentrum f{\"u}r Astronomie der Universit{\"a}t Heidelberg, M{\"o}nchhofstr. 12-14, 69120 Heidelberg, Germany\\
}

\date{Accepted XXX. Received YYY; in original form ZZZ}

\pubyear{2021}

\begin{document}
\label{firstpage}
\pagerange{\pageref{firstpage}--\pageref{lastpage}}
\maketitle

\begin{abstract}
In this paper we revisit the problem of identifying bona fide cluster Cepheids by performing an all-sky search for Cepheids associated with open clusters and making use of state-of-the-art catalogued information for both Cepheids and clusters, based on the unparalleled astrometric precision of the second and early third data releases of the Gaia satellite.
We determine membership probabilities by following a Bayesian approach using spatial and kinematic information of the potential cluster-Cepheid pairs.
We confirm 19 Cepheid-cluster associations considered in previous studies as bona-fide, and question the established cluster membership of six other associations. In addition, we identify 138 cluster Cepheid candidates of potential interest, mostly in recently discovered open clusters. 
We report on at least two new clusters possibly hosting more than one Cepheid. 
Furthermore, we explore the feasibility of using open clusters hosting Cepheids to empirically determine the Cepheid period-age relation through the use of Gaia and 2MASS photometry and a semi-automated method to derive cluster ages.
We conclude that the usage of cluster Cepheids as tentative probes of the period-age relations still faces difficulties due to the sparsely populated red giant branch and the stochastically sampled main-sequence turn-off of the open clusters, making age determinations a challenging task.
This biases the age-dateable cluster selection for Cepheid period-age studies towards older and high-mass clusters.
\end{abstract}




\section{Introduction}
\label{sec:intro}

\par Identifying Cepheid variables that are part of stellar associations and open clusters has attracted scientific attention of many astronomers, starting with \citet{Irwin1955}, and remains an important research topic. Since they are young objects, Cepheids are expected to be found in stellar associations and young open clusters. 
\par Both Cepheids and young clusters can be used to trace recent star formation events, both in the external galaxies \citep[see e.g.,][]{Payne74,Efremov2003,Glatt10}, and in the Milky Way, wherein they are expected to trace the spiral arms \citep[e.g.,][]{Magnier97a,Pietrzynksi02,Skowron19a}. Cepheids are intrinsically rare since the duration of the yellow supergiant stage for intermediate and high mass stars is short, which makes them valuable for constraining models of post-main sequence evolution \citep[e.g.,][]{Bono00b}. 
\par Cepheids in open clusters present several specific interests: first, the clusters can be used as benchmarks for abundance determinations of the Cepheids they host, which are complicated by their pulsating nature \citep[e.g.,][]{Fry1997,Lemasle17}. 
In addition, the presence of Cepheids in Galactic open clusters has proven to be extremely helpful for the calibration of the Cepheid period-luminosity relation \citep[PLR; see, e.g.,][]{Turner02,Breuval20} first found by \citet{Leavitt1912}, which makes Cepheids cornerstones of the distance scale as it provides a fundamental constraint on the Hubble constant \citep[e.g.,][]{Madore91,Riess18}.
Conversely, the existence of a PLR also allows cluster Cepheids to be used as tools to provide an independent measurement of the distance of the clusters that host them. Finally, in addition to PLRs, theoretical period-age (PA) relations have been established and can be calibrated by studying cluster-Cepheid associations. 
\par Indeed, a longer pulsation period implies a higher stellar  mass \citep{Meyer69,Kippenhahn69,Bono00a} and a younger age for the Cepheid. An empirical PA relation was derived by \citet{Tammann70} based on Galactic Cepheids and clusters, and later on
by \citet{Efremov78} using Milky Way, M31, and Large Magellanic Cloud clusters. Later, \citet{Magnier97b} obtained a new semi-empirical relation using Cepheids in the M31 star-forming region NGC~206.
Other authors have used larger samples of Cepheids and followed similar approaches to derive PA relations \citep[see,  e.g.,][]{Grebel1998,Efremov2003,Senchyna15,Inno2015}. These studies, in addition to recent theoretical approaches \citep[][]{Bono05,Anderson16,deSomma20b}, support the use of PA and PA-colour relations to supply accurate age estimates based only on a few observables. 

\par In spite of their importance, only a small number of bona fide classical Cepheids in open clusters has been reported so far. The available literature illustrates the numerous attempts to increase the list of reliable cluster-Cepheid pairs, starting with the identification of the Cepheids S~Nor and U~Sgr as members of NGC~6087 and M~25, respectively \citep{Irwin1955}, and extending throughout the last decades \citep[e.g.,][]{vanden1957,Kraft62,Efremov64,Turner86,Turner92,Turner98b,Baumgardt00,Hoyle03,Turner05,An07,Majaess08,Turner08,Turner10,Anderson13,Chen15,Lohr18,Negueruela20}. These studies have provided the approximately two tens of currently identified cluster Cepheids. They have faced common obstacles: the scarcity and the inhomogeneity of the input data. In particular, photometry originated from different sources and instruments, and accurate astrometry was rarely available. Spectroscopic information, albeit sparse,  was used by several authors when available \citep[e.g.,][]{Turner08,Anderson13,Usenko19}.

\par Large-scale astrometry-focused surveys are required to provide homogeneous catalogues for the study of the Galactic cluster Cepheid populations. The European Space Agency (ESA) mission HIgh Precision PARallax COllecting Satellite \citep[Hipparcos; ][]{ESA97} has been an invaluable source of data, allowing the  investigation of the association of many Cepheids with open clusters  \citep[see, e.g., ][]{Lynga98,Turner02}. Hipparcos' successor, the ongoing ESA Gaia mission \citep{GaiaPrusti16} revolutionized the study of stellar populations within the Milky Way. Indeed, the second and early third data releases of the Gaia catalogues (hereafter Gaia DR2 and eDR3, respectively) reach limiting magnitudes close to 21 in the $G$ band with an unparalleled astrometric precision (uncertainties $<$ 0.7\,mas for stars brighter than $G = 20$), and include parallaxes and proper motions for more than a billion stars, with photometric precisions at the millimag level \citep{GaiaBrown18,GaiaBrown20}. In particular, as far as this work is concerned, Gaia has allowed several authors to discover hundreds of new open cluster candidates and to perform membership determinations based on a full astrometric solution of the sources \citep[e.g., ][]{Ferreira19,Liu19,Torrealba19,CG19,Hunt21}. 

\par With such new data at hand, it becomes possible to update the study of Cepheids in open clusters. In this paper, we describe our investigation of Cepheid cluster membership taking advantage of the large, rich, and homogeneous Gaia data, complemented if needed by other recent surveys. We follow a Bayesian approach based on the method of \citealt{Anderson13} and using updated literature data.
Section~\ref{sec:samples} describes the data used for our study. In Sections~\ref{sec:membership} and ~\ref{sec:data}, we describe the Bayesian method used to search for cluster membership, and justify its applicability to the data characterized in Section~\ref{sec:samples}. In Section~\ref{sec:outcome} we report and discuss in detail new cluster Cepheid candidates and compare our results to previous studies. 
In Section~\ref{sec:ages} we estimate the age of our sample of cluster Cepheids by age-dating the clusters they are hosted by, and we investigate the feasibility of obtaining an empirical PA relation from our results. Finally, in Section~\ref{sec:conclusions} we summarize the contents of our study and describe their implications.

\section{Cepheids and open clusters samples}
\label{sec:samples}

\begin{figure*}
\includegraphics[angle=0,scale=.70]{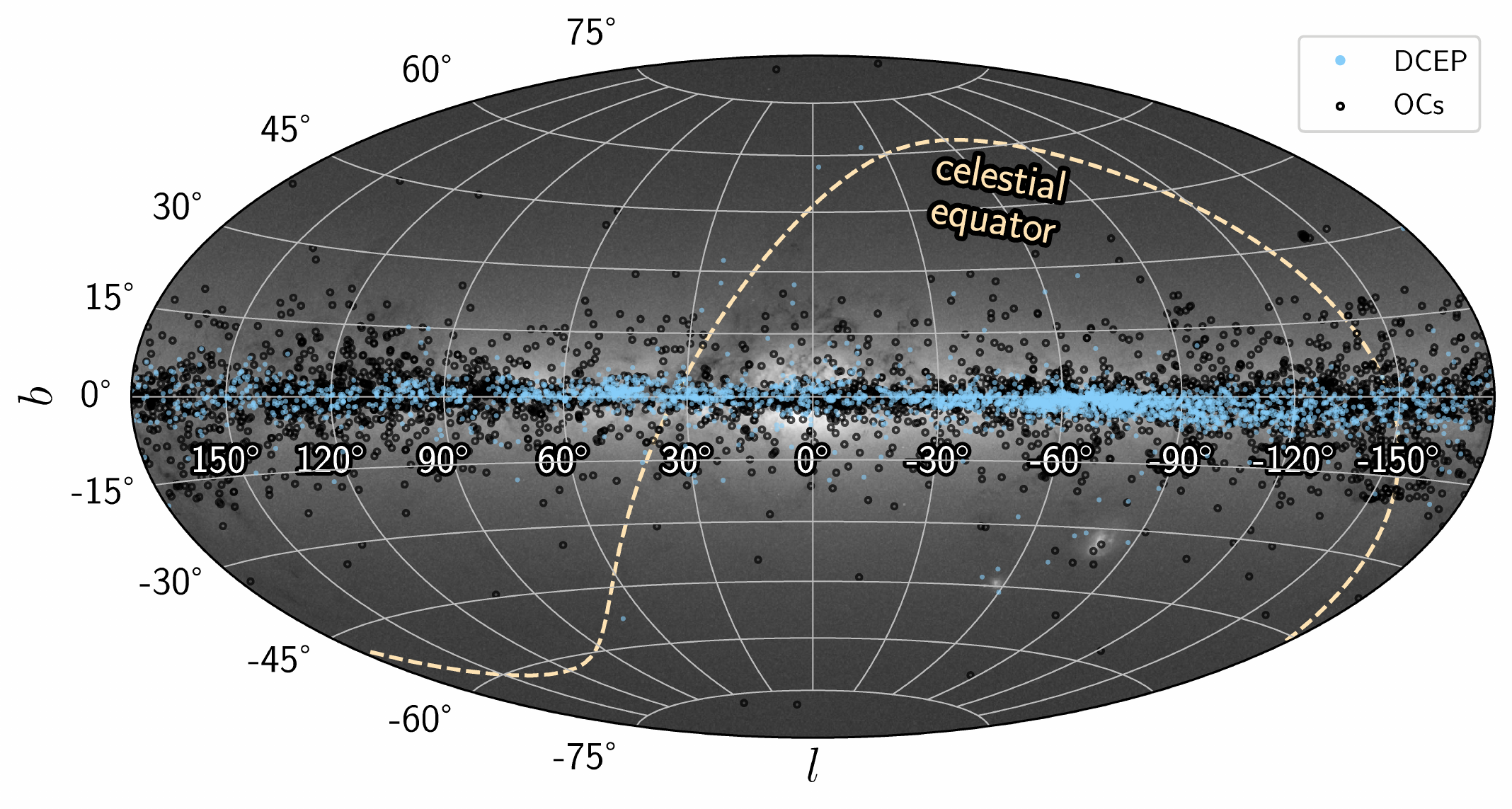}
\caption{Spatial distribution in Galactic coordinates of the Cepheids (DCEP; \textit{light blue points}) and open clusters (OCs; \textit{empty circles}) used in this work. The Gaia all-sky map in the background is shown as a reference. \textit{Image Credit: Gaia Data Processing and Analysis Consortium (DPAC); A. Moitinho / A. F. Silva / M. Barros / C. Barata, University of Lisbon, Portugal; H. Savietto, Fork Research, Portugal.}  
}
\label{fig:map}
\end{figure*}

To carry out this study, we focused on astrometric and kinematic information for both open clusters and Cepheids.

\subsection{Open clusters}
\label{sec:OCs}

We rely first on two large public catalogues of known Galactic open clusters: the compilation made by \citet[][hereafter D02]{Dias02}, for which the last update was published in 2015, and the catalogue from \citet[][hereon K13]{Kharchenko13}.
The former compilation (D02) consists of 2,167 optically visible clusters and candidates, and is based on the WEBDA database \citep{Mermilliod88}\footnote{https://webda.physics.muni.cz}, whereas the latter catalogue (K13) contains 3,006 clusters. Combining the clusters from D02 and K13 resulted in a total of 3,135 unique clusters, after removing sources classified as globular clusters by K13. We complemented those with new Milky Way open cluster or open cluster candidates from recent works based on Gaia data:

\begin{itemize}
 \item from Gaia DR2, \citet[][hereafter CG18b]{CG18b} and \citet[][henceforth CG20]{CG20b} derived the properties of $>$1200 clusters. 
 From these we include 70 and 102 clusters that do not appear in the D02/K13 catalogue, respectively;
 \item from Gaia DR2, \citet{Castro-Ginard18} identified 23 nearby open clusters (within 2\,kpc from the Sun);
 \item from Gaia DR2, \citet{Castro-Ginard19} added 53 new open clusters in the Galactic anticentre and the Perseus arm;
 \item from Gaia DR2, \citet[hereafter CG19]{CG19} added 41 clusters, most of which (33 clusters) located within 2\,kpc from the Sun;
 \item still from Gaia DR2, \citet{Ferreira19} found three clusters in the field of the intermediate-age cluster NGC~5999, and \citet{Ferreira20} discovered 25 new open cluster candidates. We added to our sample the 34 open clusters discovered by \citet{Ferreira21};
 \item from Gaia DR2, \citet{Sim19} found 207 open cluster candidates (187 totally new), all located within 1\,kpc from the Sun;
\item \citet{Liu19} identified 76 (mostly old) open cluster candidates (39 totally new) within 4\,kpc in the Gaia DR2 data
 \item the clusters found by \citet{Torrealba19} using Gaia DR2 were also added, excluding those potentially associated with the Magellanic Clouds (DES~4, DES~5, To~1, and Gaia~3);
 \item the 582 Galactic disc open clusters recently discovered by \citet{Castro-Ginard20} using again Gaia DR2 data were also incorporated in our catalogue;
 \item finally, the 41 recently discovered open cluster candidates detected by \citet{Hunt21} were also included. 

\end{itemize}

\par We gave priority to the data from D02 over K13. Both were superseded by Gaia DR2 astrometric/kinematic data from the aforementioned studies when available.
We checked for duplicated clusters by performing over this compilation an internal cross-match on the cluster center, checking individually all clusters whose centres fell within 3.5\,arcmin from each other. 
After removing evident repeated entries from this list (with astrometric parameters within one standard deviation from each other), our final catalogue contains a total of 4,140 Galactic open clusters. Their spatial distribution is shown in Figure~\ref{fig:map}. 

\subsection{Cepheids sample}
\label{sec:cepheids}

\par 
Our catalogue of classical Cepheids relies firstly on the Optical Gravitational Lensing Experiment \citep[OGLE;][]{Udalski15} database. 
We compiled the Milky Way classical Cepheids in the disc reported by \citet{Udalski18} and those in the inner disc towards the Bulge found by \citet{Soszynski2017}. We added the additional Cepheids recently added by \citet{Soszynski2020}. 
A number of additional sources list Cepheids, for instance Gaia DR2 \citep{Clementini19,Ripepi19}, the General Catalogue of Variable Stars \citep[GCVS,][]{Kukarkin69,Samus17}, the Wide-field Infrared Survey Explorer \citep[WISE,][]{Chen18}, and the Vista Variables in the V\'ia L\'actea survey \citep[VVV,][]{FerreiraLopes20}. It is notoriously difficult to classify variable stars in the near- and mid-infrared (light curves becomes more symmetric and amplitudes decrease, leading to confusion with other types of pulsating variables, eclipsing binaries and spotted stars). The purity of the infrared catalogues is then significantly lower than the purity of the optical ones \citep{Udalski18,Dekany19}. Therefore, we supplemented our catalogue with Cepheids only when they could be detected in the optical as well; hence, we did not include the 640 distant Cepheids recently discovered by \citet{Dekany19} in the Galactic midplane and bulge. 
We rely for this on the cross-survey validation performed by the OGLE team, which also includes targets from the Asteroid Terrestrial-impact Last Alert System survey \citep[ATLAS,][]{Heinze18} and the All Sky Automated Survey for Supernovae \citep[ASAS-SN,][]{Jayasinghe18,Jayasinghe19a,Jayasinghe19b} surveys. The Cepheids discovered by \citet{Clark15} in the cluster BH~222 and by \citet{Lohr18} in the clusters Berkeley~51 and Berkeley~55 entered our catalogue via this list. The extended catalogue of Cepheids has been used by \citet{Skowron19a,Skowron19b} to map the Galactic disc. Once compiled, we cross-matched this catalogue against Gaia eDR3. 
After removing Cepheids possibly related to the Magallanic Clouds (sources in the region $254^{\circ} < l < 324^{\circ}$, $-54^{\circ} < b < -22.6^{\circ}$ with negative parallaxes or with parallax-based distances $>10$\,kpc), our final sample of Milky Way classical Cepheids contains 2,921 Cepheids with Gaia eDR3 coordinates and proper motions. Their location is indicated in Figure~\ref{fig:map}.  

\section{Membership determination}
\label{sec:membership}

\par We follow the Bayesian approach adopted by \citet[][hereafter A13]{Anderson13} to address the membership determination (albeit with differences, which will be detailed in Section~\ref{diffA13}). 
The Bayesian approach enables us to quantify the likelihood of a given Cepheid being a member of an open cluster. We refer to a Cepheid-cluster pair as a ``combo", following the convention initiated by A13. To determine the membership probabilities, we use positional and kinematic constraints on both Cepheids and clusters: projected on-sky distances, parallaxes, proper motions, and radial velocities (RV) where available.

\par Using Bayes' theorem \citep[see, e.g., ][]{Jaynes03}, the posterior $P(A|B)$, i.e., the membership probability, is computed from:
\begin{equation}
P(A|B) \propto P(B|A) \times P(A)
\label{eq:bayes}
\end{equation}
\noindent where $P(B|A)$ is the conditional probability of obtaining the cluster and Cepheid data assuming that their association is real (the likelihood), and $P(A)$ (the prior) represents the probability distribution that expresses our belief of membership before the evidence used for $P(B|A)$ is considered.
We emphasize that this methodology relies on a hypothesis test that assumes membership. Thus, the association between a Cepheid and a cluster cannot be proven but only refuted by following this approach. 
\par In Equation~\ref{eq:bayes}, a normalization term associated with the probability of observing the data is neglected, given that we possess no knowledge to quantify it. In the rest of this subsection, we detail the procedure and assumptions on which the determination of the probabilities $P(B|A)$ and $P(A)$ are based.

\subsection{On-Sky Selection and Prior $P(A)$}
\label{sec:prior}

\par As other works have done in the past, we perform an initial cross-match based on the on-sky position of Cepheids and open clusters, taking into account the actual size of the clusters. The goal is to identify possible combos and to rule out Cepheids easily recognizable as non-members. This also reduces the computation time of the next steps of our analysis.

\begin{figure}
\includegraphics[angle=0,scale=.40]{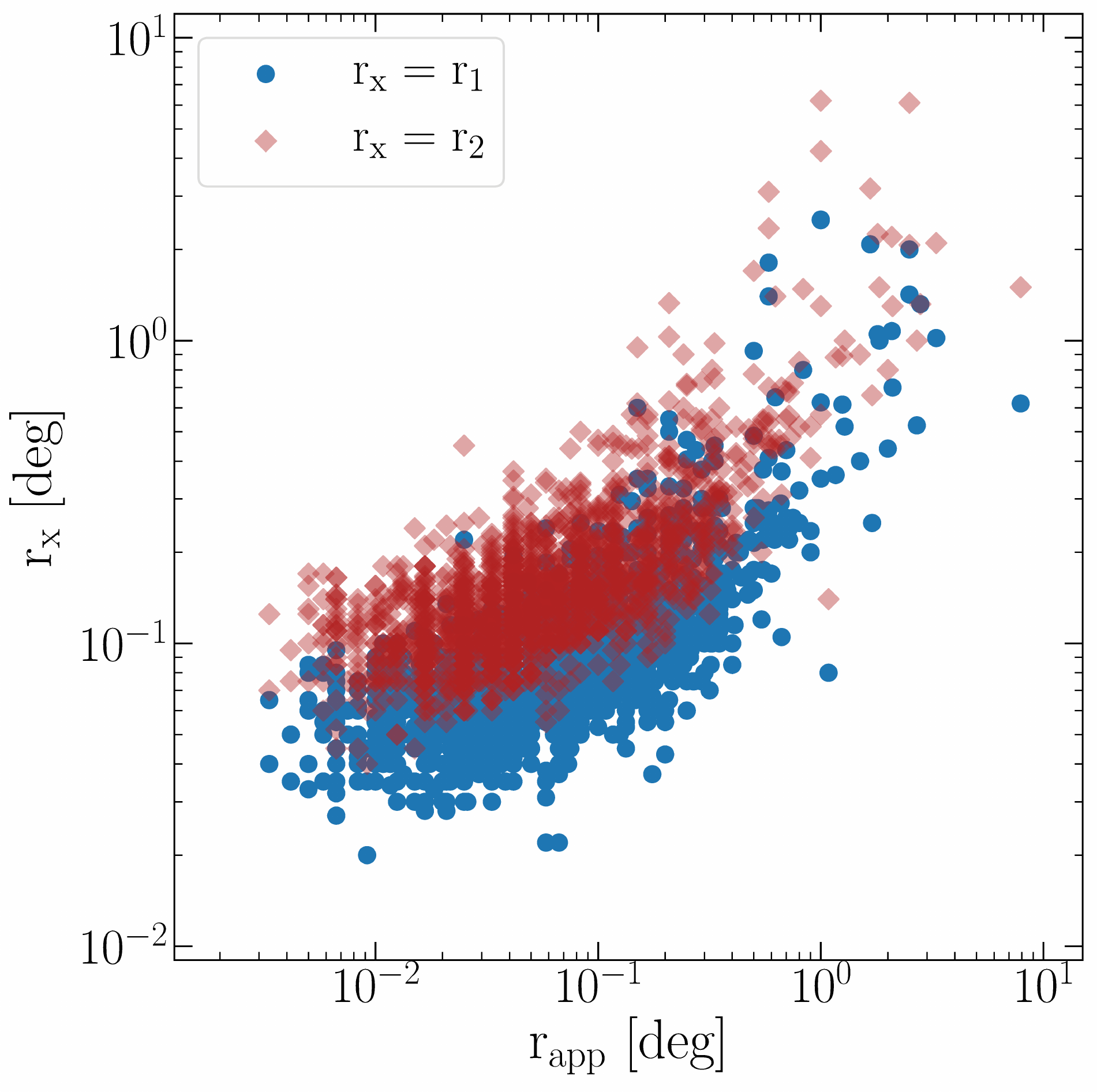}
\caption{Correlation between the open cluster apparent radii listed in D02 and the radii $r_1$ and $r_2$ presented by K13, plotted with blue dots and red diamonds, respectively.}
\label{fig:radii}
\end{figure}

\par To achieve this, we first estimate the apparent size of each cluster. K13 provide a list of cluster size parameters denoted as $r_0$, $r_1$, and $r_2$, which represent the angular radius of the core, the central part, and the entire cluster, respectively. They were fitted by eye to describe the shape of the clusters radial density profiles \citep{Kharchenko12}. Clusters from sources other than K13 do not provide these particular parameters as estimators of the cluster sizes. However, A13 demonstrated that $r_1$ and $r_2$ given by \citet{Kharchenko12} are correlated with the core and limiting radius of the clusters provided by \citet{Kharchenko05a} and \citet{Kharchenko05b}, respectively. Then, A13 used the trend between the core and limiting radii and the values from D02 to finally get an estimate of these clusters' sizes. In the case of clusters that are only present in D02, they do have an estimation of their sizes, but in the form of an angular apparent diameter. Unsurprisingly, we found correlations between the apparent radius, $r_{\rm app}$, and $r_1$ and $r_2$ (as seen in Figure~\ref{fig:radii}), but they show considerable scatter. The correlations are:
\begin{equation}
\begin{array}{lcl}
r_1 = 0.3037 \cdot r_{\rm app} + 0.0688,\\ \\
r_2 = 0.6060 \cdot r_{\rm app} + 0.1252,
\end{array}
\label{eq:r1r2}
\end{equation}

\noindent which we used to estimate $r_1$ and $r_2$ for this subsample of clusters.

\par In CG18b, CG19, and CG20, the authors present the value $r_{50}$, which represents the radius containing half the members. For these clusters, we assumed $r_1 = r_{50}$, and $r_2 = r_1 \cdot 1.957$, where the scaling factor is the ratio between $r_2$ and $r_1$ in K13, and their correlation factor is 0.962. We made the same assumptions for the clusters from \citet{Hunt21}, for which $r_{50}$ is provided. In a recent work, \citet{Sanchez20} determined the radii of a sub-sample of open clusters available in the literature based on Gaia DR2 proper motions. As a comparison, the median difference between the clusters' $r_{50}$ in CG20 and the cluster radii from \citet{Sanchez20} is 1.19\,pc for the 357 clusters in common.

\par For the clusters provided by \citet{Ferreira19,Ferreira20}, we use directly the values given in their work, as they provide estimations for both the core radius and the limiting radius.
We assumed those values to be good representations of $r_1$ and $r_2$, respectively.
\citet{Ferreira21} provide only an estimation of the clusters limiting radius $r_{\rm lim}$. Thus, for those clusters we assumed $r_2 = r_{\rm lim}$, and $r_1 = r_2/1.957$. 
Regarding the nine clusters from \citet{Torrealba19}, the authors provide a direct estimation of the half-light radius ($r_{\rm h}$). In order to obtain $r_1$ and $r_2$, we assume $r_1 = r_{\rm h}$, and applied the scaling factor 1.957 to estimate $r_2$. 

\par The case of the clusters discovered by \citet{Liu19} and \citet{Sim19} is similar.
The catalogue published by \citet{Liu19} provides the distance of the furthest member to the average member position as a proxy of the clusters' size, $r_{\rm MAX}$. We considered $r_2 = r_{\rm MAX}$ in those cases. The catalogue that characterizes the cluster candidates found by \citet{Sim19} gives the core radius as an estimation of the clusters size. We adopted those values as the clusters' $r_1$. For both the \citet{Liu19} and the \citet{Sim19} clusters, we again assumed the values of $r_2$ to be about twice as large as $r_1$.

\par For the clusters from the works by \citet{Castro-Ginard18,Castro-Ginard19,Castro-Ginard20}, a list of tentative cluster members was also published by the authors. 
With this information we computed the radii $r_1$ and $r_2$ using the previously mentioned scaling factor and setting $r_1 = r_{50}$, the median cluster member distance \citep[as provided by][]{Castro-Ginard18,Castro-Ginard19,Castro-Ginard20} with respect to their tabulated central coordinates. \\

\par With an estimated cluster size, it is possible to perform the initial cross-match based on the on-sky position of the Cepheid-open cluster pairs. 
To minimize the list of possible combinations for which we compute probabilities, we selected all Cepheids within the largest distance between 2\,deg from a given cluster centre, or five times the cluster's $r_1$. This is similar to the procedure followed by A13. With the median $r_1$ of the entire sample at 0.08\,deg and with $\sim 95$\,per cent of the clusters having values of $r_1 < 0.4$\,deg, we consider that finding a Cepheid at more than $2$\,deg from the cluster centre makes its membership unlikely and justifies such an arbitrary choice. For the remaining clusters, for which the median $r_1$ is 0.6\,deg, choosing the limit at 5$\cdot r_1$ allows for some flexibility, even when computing probabilities for nearby associations with $r_1 > 1$\,deg, such as Collinder~285. 

\par A total of $\sim$ 44,300 possible combos results from this procedure. 
We focus on these combinations in the next steps of our membership analysis, starting with the determination of the prior.\\

\par We define the prior $P(A)$ following A13's approach: we only take into account the on-sky separation between a Cepheid and a cluster, and the apparent size of the latter based on its core and limiting radius. By defining the quantity $x$ as:
\begin{equation}
x = \frac{r - r_1}{2\cdot r_2 - r_1}\\
\label{eq:x}
\end{equation}
where $r$ is the on-sky separation, $r_{1}$ is a proxy for the core radius, and $r_{2}$ a proxy for the limiting radius, we can measure the relative position of the Cepheid with respect to the centre of the cluster, weighted by its size. Then, for the value of the probability we define:
\begin{equation}
\begin{array}{ll}
P(A) \equiv 1, & x < 0\\ \\
P(A) \equiv 10^{-x}, & x \geq 0
\end{array}
\label{eq:PA}
\end{equation}

\par From this definition, a combo's prior probability will be $1$ if the Cepheid falls within the core of the cluster, and will reach $10$\,per cent at $x=1$, inspired by the exponential decline of the radial profile of star clusters. We note that in the study by A13 the authors define the prior such that it reaches $0.1$\,per cent at $x=1$. Considering only the prior, this means that we are more flexible than A13 when computing probabilities.

\par We modified the prior with respect to A13 to take into account the recent results of e.g.,  \citet{Meingast21}: from a sample of young open clusters with ages between 30 and 300~Myr (perfectly matching the ages of Cepheids), they found that almost all clusters are surrounded by a large halo a stars they call coronae and which extend further than 100~pc from the clusters' centre. Most clusters show evidences of expansion along one or more spatial axes, a feature also observed by, e.g., \citet{Pang20}. Although reminiscent of tidal features observed in older clusters, such features are most likely primordial and related to filamentary star formation \citep[e.g.,][]{Beccari20,Tian20}.
With a somewhat relaxed prior, we are more likely to recover bona-fide cluster Cepheids located at large distances from the cluster centre, but also more exposed to spurious detections. We note in passing that for a Cepheid located at 5$\cdot r_1$ from the centre of a cluster (one of the pre-selection criteria adopted above), Equation~\ref{eq:PA} would return $P(A)\approx$ 0.05 when $r_2\approx 2\cdot r_1$. 

\subsection{The Likelihood $P(B|A)$}
\label{sec:likelihood}

\par As other authors have done in the past (\citealt{Robichon99,Baumgardt00}; A13; \citealt{Hanke20,Prudil21}), we determine the likelihood of membership $P(B|A)$ as a hypothesis test based on the Mahalanobis distance \citep[introduced by][]{Mahalanobis36}, which is a measure of the distance between a point (a Cepheid) and a distribution (an open cluster). 
The task of calculating quantiles for multivariate normal distributions is not as simple as in the one-dimensional case, since these quantiles can be considered ellipsoids in dimensions higher than two. However, calculating the Mahalanobis distance is a rather simple method to describe all points on the surface of such a multidimensional ellipsoid. 
\par Given a vector $z$ built as the difference $\Delta$ between the Cepheid and the mean cluster parameters (here we consider the parallaxes $\varpi$, proper motion in right ascension $\mu^*_\alpha$ and declination $\mu_\delta$, and the radial velocities $V_r$): 
\begin{equation}
\vec{z} = ( \Delta \varpi, \ \Delta \mu^*_\alpha, \ \Delta \mu_\delta, \ \Delta V_r )\\
\label{eq:z}
\end{equation}
\noindent the square of the Mahalanobis distance between the Cepheid and the cluster, $c$, can be expressed as:
\begin{equation}
c = \vec{z}^{\ T}\ \Sigma^{-1}\ \vec{z},\\
\label{eq:c}
\end{equation}
\noindent where $\vec{z}^{\ T}$ is the transpose of $\vec{z}$, and $\Sigma^{-1}$ denotes the inverse of the sum of the covariance matrices of a cluster and a Cepheid when systematic effects and correlations are taken into account. It should be noted that, for the purpose of these calculations, we compute $c$ under the assumption that the Cepheid was not used to measure the mean cluster parameters.
\par An additional quantity that we used to determine $\Sigma^{-1}$ is the re-normalized unit weight error (RUWE), which is given as a parameter in the Gaia DR2 and eDR3 catalogues and which accounts for the fitting effects when the astrometric solution is poor \citep{Lindegren18}. For sources where the Gaia single-star model is well fitted, the RUWE is expected to be close to 1.0, and values significantly greater than 1.0 reflect problems in the astrometric solution or non-single objects. To account for this, for stars with RUWE $> 1.4$ we scaled the elements of the covariance matrix that are taken from the Gaia catalogues by the square of their RUWE values. If a given Cepheid has a non-numerical RUWE, we set its value to $22$, which corresponds to the maximum value of the original list of Cepheids in the Gaia catalogue.

\par We assume the clusters' covariance matrices to be diagonal whenever the data collected come from several sources, and we possess no information about possible correlations. If correlations between the cluster parameters are known and given in the cluster sources or can be inferred from, for example, Gaia data, the mean values from the clusters members are included in the corresponding cluster's covariance matrix.
For the Cepheids' matrices we used the correlations between parallaxes and proper motions explicitly provided by the Gaia eDR3 catalogue.
Finally, we assume the correlation between the properties of clusters and Cepheids to be negligible. 

\par It is possible to show that, under these conditions, and assuming Gaussian distributions, the Mahalanobis distance 
$c$ is actually $\chi^2$-distributed. It is worth mentioning that the shape of the $\chi^2$ distribution depends on the number of dimensions of $\vec{z}$, i.e., the number of constraints considered for the combo. Finally, the likelihood is the result of:
\begin{equation}
P(B|A) = 1 - p(c)\\
\label{eq:pba}
\end{equation}
\noindent where $p(c)$ represents the probability of finding a value at least as extreme as the observed $c$ under the null hypothesis of (true) membership, that is, is the p-value of $c$.

\subsection{Differences with A13}
\label{diffA13}

\par Although our study follows the Bayesian approach of A13, there are also significant differences, both in the method and in the data, since we benefit from greatly improved data quality thanks to Gaia in particular. We discuss these differences here below:  

\par A13 initially considered 2,168 clusters. In our study, many more entered the catalogue, since the D02 database has been continuously updated and thanks to the discovery of numerous open clusters after Gaia DR2. Our catalogue contains over 4,000 clusters. Moreover, cluster parallaxes and proper motions have been updated to Gaia DR2 values for roughly half of our cluster catalogue.

\par Our initial catalogue of Cepheids is also much larger than the one used in A13 (2,921 vs. 1,821) thanks to numerous surveys having provided a large number of new Cepheids. Parallaxes in A13 are taken directly from the Hubble Space Telescope \citep[8 stars, ][]{Benedict07}, from the study by \citet{Storm11a} for 33 stars, and from Hipparcos \citep{vanLeeuwen07} for a good fraction of their sample. For 622 Cepheids, parallaxes are derived by inverting the distance to the Cepheid, computed using a period-luminosity relation in the $V$-band since the largest photometric data sets are available in this band. Although computed with the greatest care, this method suffers from the intrinsic width of the instability strip, which is much larger than in the near-infrared for instance, from the metallicity dependence of $V$-band period-luminosity relations \citep[e.g.,][]{Romaniello2008}, from the heterogeneity of the photometric data sets used, and from very large uncertainties on the adopted values for $E(B-V)$. We benefit instead from the great quality of Gaia eDR3 parallaxes. Finally, the proper motions for Cepheids in A13 are taken from Hipparcos \citep{vanLeeuwen07} when available and from the Position and Proper Motion Extended-L catalogue \citep[PPMXL;][]{Roeser10} otherwise, while ours are also from Gaia eDR3. Radial velocities are comparable in terms of data availability, accuracy, and precision.

\par
\begin{figure*}
\begin{center}
\includegraphics[angle=0,scale=.45]{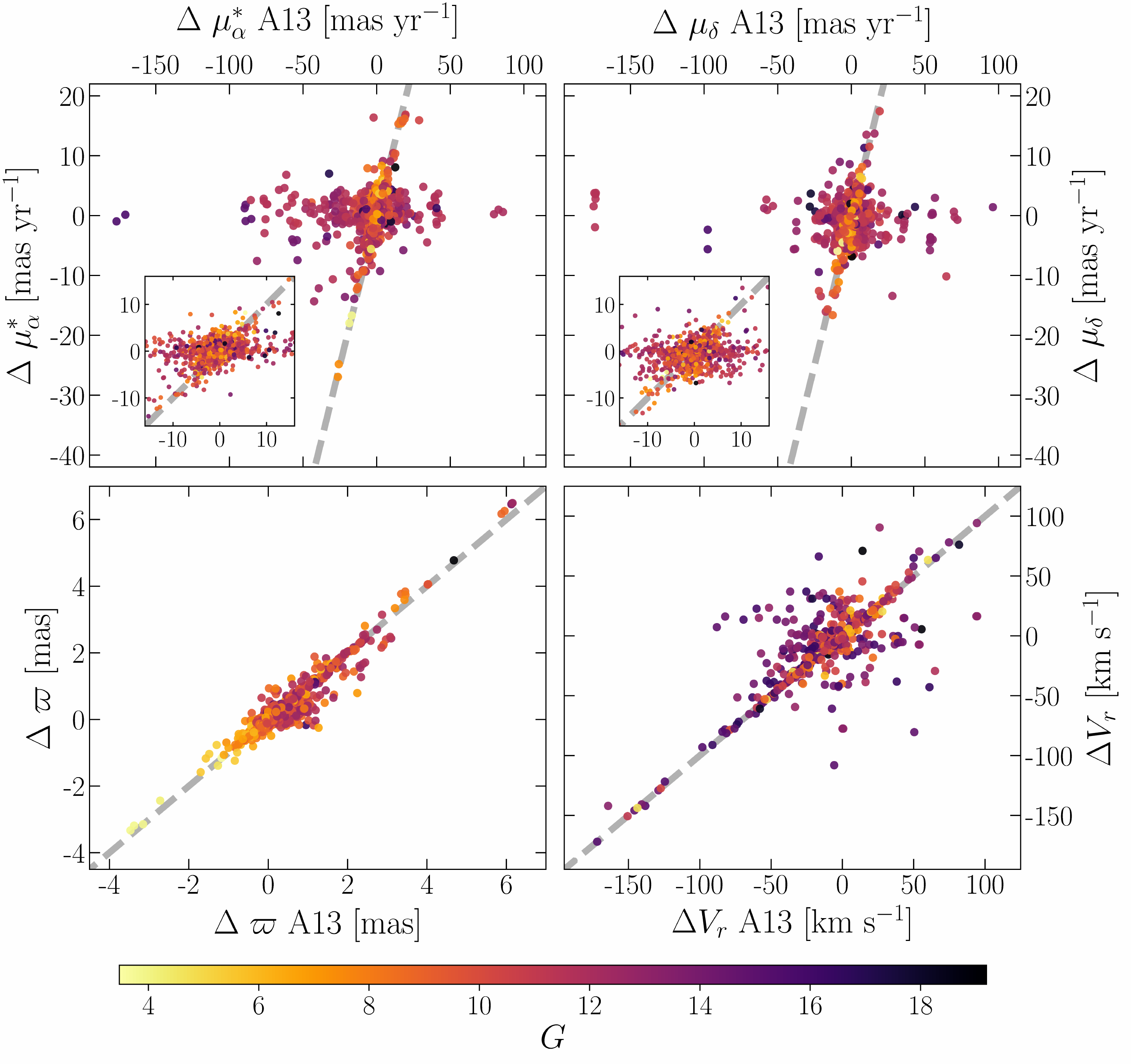}
\end{center}
\caption{Comparison of the differences in proper motions, parallaxes, and radial velocities between clusters and Cepheids for the combos in common with A13. The filled circles shown are colour-coded by the Cepheids' mean G magnitudes. 
A grey dashed line representing the identity function is plotted in each panel as a reference. 
In the plots displaying proper motion differences (\textit{upper panels}) we include an enlargement of the central distribution of points.}
\label{fig:A13comparison}
\end{figure*}

 In Figure~\ref{fig:A13comparison}, we compare the astrometric and kinematic data sets, after cross-matching the $\sim$4,000 pairs investigated by A13 against our $\sim$44,300 candidates. Since no other comparison data are available, we compare the difference between a given Cepheid and its potential host in both data sets. On the other hand, such a comparison relates directly to possible differences in the probabilites $P(B|A)$ derived by each study. We note that even small discrepancies may be significant in the computation of $P(B|A)$. In terms of proper motions, the spreads of the distributions are of the order of tens of \,mas\ yr$^{-1}$, and can be as high as $\sim$180\,mas\ yr$^{-1}$.

\par We mentioned earlier that we adopted a different prior for cluster membership, to take into account the mounting evidence of large spatial extensions even for young clusters. Since this particular property had not been discovered yet, A13 adopted a stricter prior that is less sensitive to the clusters' outskirts but conversely less prone to false positives.
We note in passing that an even looser prior has been used by \citet{Hanke20} in a search for extra-tidal halo stars in the neighborhood of globular clusters.

\par Another important difference is that, in contrast to A13, who took into account up to six dimensions when computing the square of the Mahalanobis distance, we only consider up to four ($\varpi$, $\mu^*_\alpha$, $\mu_\delta$, and $V_r$), and neither ages nor metallicities, in the computation of $c$.

\par If we were to include ages in our analysis, we would necessarily have computed the Cepheids' ages from period-age relations. This is, however, impossible in our case since one of our goals is to constrain such relations. Moreover, ages derived from period-age relations with an average stellar rotation ($\Omega_{\rm ZAMS}/\Omega_{\rm crit}=0.5$, \citealt{Anderson16}) are 50-100\,per cent higher than those derived from period-age relations without rotation \citep[e.g.,][]{Bono05}. As far as clusters are concerned, for 93\,per cent of the clusters in D02 ages were provided. From the numerous clusters discovered in Gaia DR2 data, a good fraction has ages available \citep[][CG20, without quoted uncertainties for the vast majority]{Bossini19}. However, the difficulty to properly identify the main-sequence turnoff (MSTO) compromises the age determination via the isochrone-fitting method for young clusters, that is, those potentially hosting Cepheids, especially when using an automated algorithm as we experienced ourselves (see Section~\ref{sec:ages} for a more detailed discussion). The difficulty is reduced towards higher ages, when the MSTO and the red giant branch (RGB) become more populated, and A13 allowed for varying uncertainties in age to take this effect into account. As pointed out by A13, even large uncertainties are useful, in the sense that they allow us to filter out pairs for which ages definitely mismatch. Our concern here is that pairs could be rejected on the basis of an inaccurate age determination, even when allowing for large uncertainties.

\par The situation is even worse in the case of the metallicity, since [Fe/H] is available only for a small fraction of the clusters, and often relies on disparate techniques (photometric estimates, low-resolution spectroscopy, high-resolution spectroscopy) and very few cluster members. Many have attempted to provide homogeneous metallicity scales in the recent past, either within a given large spectroscopic survey or by collecting data from various sources \citep[e.g.,][]{Netopil16}, but the number of clusters with available metallicities falling in the age range considered here is very small. Studies of the Milky Way radial abundance gradient indicate a good agreement between $<$ 1~Gyr old open clusters and Cepheids, both for [Fe/H] and other elements \citep[e.g., ][]{Lemasle08,Genovali15,Magrini17}, but detailed comparisons are still missing.
\cite{Fry1997} found a good agreement ($\approx$ 0.1~dex) between the metallicity of two main-sequence stars and the Cepheid U~Sgr in M25, but the comparison is limited to [Fe/H]. \citet{Lemasle17} found a similarly good agreement between 6 Cepheids and a large number of RGB stars analyzed by \citet{Mucciarelli11} in the young LMC cluster NGC~1866, which contains 24 Cepheids \citep{Musella16} and is therefore not representative of Milky Way young clusters.

\par On the basis of the above, we decided to search for possible cluster-Cepheid combos relying on astrometry and kinematics only, awaiting for a membership confirmation from studies including a detailed age and chemical analysis.

\section{The data}
\label{sec:data}

\subsection{Parallaxes}
\label{sec:plx}

\par For the open clusters in our list (see Section~\ref{sec:samples}) we used the parallaxes and their associated uncertainties directly from their original sources, if available. If not (as for the D02 clusters), we derived the parallaxes $\varpi$ from the published distances $d$ in parsecs, following: 
\begin{equation}
\begin{array}{lcl}
\varpi = \frac{1000}{d} \ {\rm (mas)}, \\ \\
\sigma_{\varpi} = \frac{1000}{d^2} \cdot \sigma_d \ {\rm (mas)}, 
\end{array}
\label{eq:parallax}
\end{equation}
\noindent where $\sigma_{d}$ represents the distance uncertainty, and $\sigma_{\varpi}$ the corresponding assumed parallax uncertainty. 
We consider this approximation justified since there is an overall good agreement between parallaxes computed this way and those provided by CG20 using Gaia DR2 data with a median difference of 0.08\,mas for the clusters in common between that study and D02. In such a case we enforced a distance uncertainty of $20$\,per cent (as done by A13) to account for various effects impacting the distance determination such as stellar rotation and binarity. 

\par In any case, older values are superseded by parallaxes and parallax errors from Gaia DR2 when available. Moreover, \citet{Lindegren18} reported a (global) zero-point shift of $-0.03$\,mas for Gaia DR2 parallaxes. Unless this was explicitly accounted for in the original papers, we shifted the Gaia DR2 parallaxes accordingly. 

\par In addition, parallaxes of Gaia sources located close to each other on the sky are highly correlated, especially when they are separated by less than one degree. Following the recommendations of \citet{Lindegren18} for sources separated by an angle $\theta$, we computed the mean spatial covariance of the parallax errors $V_{\varpi}(\theta)$ of the members of each cluster in our sample based on Gaia DR2 data, if a list of members was available in the source catalogue, and if these spatial covariances were not originally considered. Otherwise, spatial covariances were neglected. The corresponding systematic uncertainties and additional correlations are then included in the covariance matrices of the clusters. 
\par For the Cepheids, parallaxes from Gaia eDR3 are available for all the stars in our sample. We corrected them for the Gaia eDR3 zero-point parallax offset following \citet{Lindegren20}\footnote{https://gitlab.com/icc-ub/public/gaiadr3\_zeropoint}, and we increased the uncertainties by 10\,per cent to account for their likely underestimation, based on the work of \citet{Fabricius20}.  
As in the case of the clusters, these changes were included in the covariance matrices of the Cepheids. 

\subsection{Proper Motions}
\label{sec:pms}

\par The (mean) proper motions in right ascension and declination  and their respective uncertainties were first taken from D02. Some clusters are registered only in K13, where a single value $\sigma_{\mu}$ is given for the proper motion uncertainties. We therefore adopted $\sigma_{\mu_\alpha}^{*} = \sigma_{\mu_\delta} = \sigma_{\mu}/\sqrt{2}$. These values were replaced by proper motions based on Gaia DR2 when available. In this case, we adopted the values tabulated in the respective source catalogues for the proper motions and their uncertainties.

Similarly to the $V_{\varpi}(\theta)$ correction described above, we corrected the Gaia DR2 proper motion uncertainties by taking into account the spatial covariances $V_{\mu}(\theta)$ that affect sources located close to each other in the sky \citep{Lindegren18}, in this case the members of a given cluster, and added the mean $V_{\mu}(\theta)$ of such a cluster as a systematic uncertainty when this effect was not already accounted for in the original studies. 
For the clusters with Gaia DR2 proper motions but no list of members (which have sizes as small as $\sim$ 0.01\,deg), we added a systematic uncertainty of 0.066\,mas\ yr$^{-1}$ following the recommendations of \citet{Vasiliev19} (see their Figure~3). 
These changes were propagated in the cluster covariance matrices.  

\par In the case of the Cepheids, we take their proper motions directly from the Gaia eDR3 catalogue. We revised these values using the Gaia eDR3 proper motion bias correction recently described by \citet{CG21}, and added a systematic error of -10\,$\mu$as\,yr$^{-1}$ to the uncertainties of the Cepheids with $G<$ 13 listed in the catalogue to account for the remaining color-dependent systematics discussed by these authors. These changes were included in the covariance matrices of the Cepheids. 

\subsection{Radial Velocities}
\label{sec:rvs}

\par For the radial velocity of the open clusters, we use the mean values and uncertainties listed in D02. We note, however, that for a given cluster, the uncertainty on the radial velocity provides a good estimate of the intrinsic velocity dispersion only when the number of stars analyzed is large. For clusters only present in K13, where no uncertainty is given, we follow A13's reasoning and estimate this value by computing $\sigma_{ocRV} = 10 / \sqrt{N_{RV}}$\,km\ s$^{-1}$, where $N_{RV}$ is the number of stars used to compute the radial velocity of the cluster. 
If $N_{RV}$ is also missing, we assume a value of $15$\,km\ s$^{-1}$, which matches the maximum velocity dispersion for open clusters recently analyzed by, e.g., \citet{Carrera19} and \citet{Donor20}.

\par To update the radial velocities and associated uncertainties of our cluster catalogue we used the information of 131 clusters listed in \citet{Carrera19}, obtained by data mining the Apache Point Observatory Galactic Evolution Experiment \cite[APOGEE DR14,][]{Abolfathi18,Holtzman18} and the Galactic Archaeology with HERMES survey \cite[GALAH DR2,][]{Buder18}. 
Of these open clusters, 127 are in common with the original D02 + K13 catalogs, of which 72 have previously derived radial velocities (in D02 + K13).
We also considered the radial velocities of the 128 clusters listed in \citet{Donor20}, who examined APOGEE data \citep[DR16,][]{Joensson20}.
Of these clusters, 126 are listed in the catalogue of D02 and K13.
We note that the mean absolute radial velocity difference between the values listed in the D02 and K13 catalogues, and those catalogued by \citet{Carrera19} and \citet{Donor20} is $12.1 \pm 26.1$\,km\ s$^{-1}$ and $7.5 \pm 13.9$\,km\ s$^{-1}$, respectively.
For the clusters for which more than one radial velocity measurement is available, we give priority to the more recent studies over older ones (including those in D02 and K13, since we favour the homogeneity of the data). 
In any case, we impose a minimum velocity dispersion of $2$\,km\ s$^{-1}$, a value also in line with the compilations of \citet{Carrera19} and \citet{Donor20} for clusters in which the measured velocity dispersion relies on more than 10 stars.

\par As classical Cepheids are pulsating stars, monitoring observations are required to derive their systemic velocity (or alternatively radial velocity templates). Such data are in general not available, and we rely on the compilation made by \citet{Melnik15}, in which they provide heliocentric radial velocities and their uncertainties for $\sim$ 320 Cepheids from different sources (see references in \citealt{Melnik15}). To take into account possible phase coverage biases or binarity \citep[up to 80\,per cent of Cepheids are in binary systems,][]{Kervella19a}, we impose an arbitrary minimum uncertainty for the Cepheids' radial velocities of $2$\,km\ s$^{-1}$.

\begin{table*}
\caption{Cluster - Cepheid pairs with membership probabilities $P(A|B)$~$>$ 0.10.  The table lists the cluster names as well as their MWSC identification in the K13 catalogue, the Cepheid names, the ratio between the separation of the pair and the cluster's $r_1$ ($\rm{Sep}/r_1$), the list of constraints used to derive the membership probability, the prior $P(A)$, the likelihood $P(B|A)$ and the membership probability $P(A|B)$.}
\centering
\label{tab:combos}
{\scriptsize
\begin{tabular}{cccHcccccH}  
\hline
Open cluster & MWSC ID & Cepheid & RUWE & $\rm{Sep}/r_1$ & Constraints & $P(A)$ & $P(B|A)$ & $P(A|B)$ & Ref.\\
\noalign{\smallskip}
\hline
\noalign{\smallskip}
   Trumpler~14 &  1846 &              OGLE-GD-CEP-1673 &   1.03 &   0.57 &         $\varpi$, $\mu_{\alpha}^{*}$, $\mu_{\delta}$ &  1.00 &  1.00 &  1.00 &  -- \\
       UBC~553 &    -- &              OGLE-GD-CEP-1194 &   1.02 &   0.54 &         $\varpi$, $\mu_{\alpha}^{*}$, $\mu_{\delta}$ &  1.00 &  0.99 &  0.99 &  -- \\
   Berkeley~55 &  3490 &  ASASSN-V~J211659.90+514558.7 &   1.09 &   0.29 &         $\varpi$, $\mu_{\alpha}^{*}$, $\mu_{\delta}$ &  1.00 &  0.94 &  0.94 &  -- \\
        Gaia~5 &    -- &                     V0423~CMa &   1.15 &   0.81 &                                             $\varpi$ &  1.00 &  0.94 &  0.94 &  -- \\
       ASCC~79 &  2288 &          OGLE-GD-CEP-1752$^*$ &   1.02 &   0.78 &         $\varpi$, $\mu_{\alpha}^{*}$, $\mu_{\delta}$ &  1.00 &  0.94 &  0.94 &  -- \\
   Berkeley~51 &  3280 &  ASASSN-V~J201151.18+342447.2 &   1.07 &   0.75 &         $\varpi$, $\mu_{\alpha}^{*}$, $\mu_{\delta}$ &  1.00 &  0.85 &  0.85 &  -- \\
    Harvard~16 &  2616 &              OGLE-BLG-CEP-041 &   5.02 &   0.75 &         $\varpi$, $\mu_{\alpha}^{*}$, $\mu_{\delta}$ &  1.00 &  0.83 &  0.83 &  -- \\
      FSR~0951 &   849 &                        RS~Ori &   1.12 &   0.24 &  $\varpi$, $V_r$, $\mu_{\alpha}^{*}$, $\mu_{\delta}$ &  1.00 &  0.82 &  0.82 &  -- \\
       Lynga~6 &  2348 &                        TW~Nor &   0.89 &   0.39 &  $\varpi$, $V_r$, $\mu_{\alpha}^{*}$, $\mu_{\delta}$ &  1.00 &  0.82 &  0.82 &  -- \\
      NGC~7790 &  3781 &                        CF~Cas &   1.04 &   0.30 &  $\varpi$, $V_r$, $\mu_{\alpha}^{*}$, $\mu_{\delta}$ &  1.00 &  0.80 &  0.80 &  -- \\
    Gulliver~9 &    -- &                        AM~Vel &  41.08 &   1.33 &         $\varpi$, $\mu_{\alpha}^{*}$, $\mu_{\delta}$ &  0.77 &  1.00 &  0.77 &  -- \\
       IC~4725 &  2940 &                         U~Sgr &   0.85 &   0.10 &  $\varpi$, $V_r$, $\mu_{\alpha}^{*}$, $\mu_{\delta}$ &  1.00 &  0.75 &  0.75 &  -- \\
       NGC~129 &    53 &                        DL~Cas &   1.88 &   0.04 &  $\varpi$, $V_r$, $\mu_{\alpha}^{*}$, $\mu_{\delta}$ &  1.00 &  0.75 &  0.75 &  -- \\
    Czernik~41 &  3192 &             J297.7863+25.3136 &   0.99 &   0.65 &         $\varpi$, $\mu_{\alpha}^{*}$, $\mu_{\delta}$ &  1.00 &  0.73 &  0.73 &  -- \\
     vdBergh~1 &   934 &                        CV~Mon &   1.10 &   0.68 &         $\varpi$, $\mu_{\alpha}^{*}$, $\mu_{\delta}$ &  1.00 &  0.67 &  0.67 &  -- \\
      NGC~6193 &  2444 &          OGLE-GD-CEP-1175$^*$ &   1.00 &   0.70 &         $\varpi$, $\mu_{\alpha}^{*}$, $\mu_{\delta}$ &  1.00 &  0.67 &  0.67 &  -- \\
      NGC~6067 &  2370 &                     V0340~Nor &   0.92 &   0.20 &         $\varpi$, $\mu_{\alpha}^{*}$, $\mu_{\delta}$ &  1.00 &  0.66 &  0.66 &  -- \\
        BH~222 &  2564 &              OGLE-BLG-CEP-110 &   0.98 &   0.41 &         $\varpi$, $\mu_{\alpha}^{*}$, $\mu_{\delta}$ &  1.00 &  0.65 &  0.65 &  -- \\
      NGC~6649 &  2949 &                     V0367~Sct &   1.02 &   0.73 &         $\varpi$, $\mu_{\alpha}^{*}$, $\mu_{\delta}$ &  1.00 &  0.63 &  0.63 &  -- \\
 Kronberger~84 &  3532 &  ASASSN-V~J213533.70+533049.3 &   0.95 &   0.22 &         $\varpi$, $\mu_{\alpha}^{*}$, $\mu_{\delta}$ &  1.00 &  0.62 &  0.62 &  -- \\
       UBC~266 &    -- &              OGLE-GD-CEP-1676 &   1.06 &   1.08 &         $\varpi$, $\mu_{\alpha}^{*}$, $\mu_{\delta}$ &  0.94 &  0.62 &  0.58 &  -- \\
      FSR~1755 &    -- &              OGLE-BLG-CEP-175 &   1.09 &   1.00 &                   $\mu_{\alpha}^{*}$, $\mu_{\delta}$ &  1.00 &  0.57 &  0.57 &  -- \\
       UBC~130 &    -- &                        SV~Vul &   1.20 &   1.41 &         $\varpi$, $\mu_{\alpha}^{*}$, $\mu_{\delta}$ &  0.73 &  0.71 &  0.52 &  -- \\
       UBC~229 &    -- &                     V0335~Pup &   1.05 &   0.48 &         $\varpi$, $\mu_{\alpha}^{*}$, $\mu_{\delta}$ &  1.00 &  0.51 &  0.51 &  -- \\
      NGC~7790 &  3781 &                      CE~Cas~B &   1.20 &   0.51 &         $\varpi$, $\mu_{\alpha}^{*}$, $\mu_{\delta}$ &  1.00 &  0.50 &  0.50 &  -- \\
      FSR~0158 &  3182 &                        GQ~Vul &   0.91 &   1.80 &         $\varpi$, $\mu_{\alpha}^{*}$, $\mu_{\delta}$ &  0.53 &  0.89 &  0.47 &  -- \\
       ASCC~12 &   427 &                        SV~Per &  11.69 &   1.85 &         $\varpi$, $\mu_{\alpha}^{*}$, $\mu_{\delta}$ &  0.51 &  0.90 &  0.46 &  -- \\
       LP~1937 &    -- &                        DF~Cas &   1.05 &   1.58 &         $\varpi$, $\mu_{\alpha}^{*}$, $\mu_{\delta}$ &  0.64 &  0.71 &  0.45 &  -- \\
       UBC~608 &    -- &  ASASSN-V~J040516.13+555512.9 &   1.06 &   0.63 &         $\varpi$, $\mu_{\alpha}^{*}$, $\mu_{\delta}$ &  1.00 &  0.45 &  0.45 &  -- \\
       LP~1370 &    -- &                        DT~Gem &   1.19 &   1.77 &         $\varpi$, $\mu_{\alpha}^{*}$, $\mu_{\delta}$ &  0.55 &  0.78 &  0.43 &  -- \\
      FSR~0172 &  3218 &                    Dauban~V16 &   0.96 &   1.49 &         $\varpi$, $\mu_{\alpha}^{*}$, $\mu_{\delta}$ &  0.68 &  0.59 &  0.40 &  -- \\
      NGC~6087 &  2382 &                         S~Nor &   0.88 &   0.14 &  $\varpi$, $V_r$, $\mu_{\alpha}^{*}$, $\mu_{\delta}$ &  1.00 &  0.38 &  0.38 &  -- \\
       LP~2134 &    -- &                        VY~Per &   1.15 &   1.53 &         $\varpi$, $\mu_{\alpha}^{*}$, $\mu_{\delta}$ &  0.66 &  0.56 &  0.37 &  -- \\
        LP~888 &    -- &                      CE~Cas~B &   1.20 &   1.91 &         $\varpi$, $\mu_{\alpha}^{*}$, $\mu_{\delta}$ &  0.50 &  0.71 &  0.35 &  -- \\
       LP~2134 &    -- &                        UY~Per &   1.17 &   1.01 &         $\varpi$, $\mu_{\alpha}^{*}$, $\mu_{\delta}$ &  0.99 &  0.35 &  0.35 &  -- \\
      NGC~6631 &  2916 &              OGLE-BLG-CEP-164 &   1.20 &   1.43 &         $\varpi$, $\mu_{\alpha}^{*}$, $\mu_{\delta}$ &  0.71 &  0.46 &  0.33 &  -- \\
        LP~888 &    -- &                      CE~Cas~A &   1.09 &   1.91 &         $\varpi$, $\mu_{\alpha}^{*}$, $\mu_{\delta}$ &  0.50 &  0.66 &  0.33 &  -- \\
       UBC~106 &    -- &                        CM~Sct &   0.84 &   1.30 &         $\varpi$, $\mu_{\alpha}^{*}$, $\mu_{\delta}$ &  0.79 &  0.40 &  0.32 &  -- \\
      DBSB~179 &  2544 &              OGLE-BLG-CEP-173 &   0.98 &   0.66 &         $\varpi$, $\mu_{\alpha}^{*}$, $\mu_{\delta}$ &  1.00 &  0.31 &  0.31 &  -- \\
       IC~2395 &  1537 &          OGLE-GD-CEP-0270$^*$ &   1.07 &   1.93 &         $\varpi$, $\mu_{\alpha}^{*}$, $\mu_{\delta}$ &  0.48 &  0.61 &  0.29 &  -- \\
        BH~121 &  1960 &              OGLE-GD-CEP-1688 &   1.00 &   0.91 &         $\varpi$, $\mu_{\alpha}^{*}$, $\mu_{\delta}$ &  1.00 &  0.29 &  0.29 &  -- \\
       UBC~291 &    -- &              OGLE-GD-CEP-1719 &   1.20 &   2.21 &         $\varpi$, $\mu_{\alpha}^{*}$, $\mu_{\delta}$ &  0.40 &  0.72 &  0.28 &  -- \\
        LP~888 &    -- &                        CF~Cas &   1.04 &   1.98 &         $\varpi$, $\mu_{\alpha}^{*}$, $\mu_{\delta}$ &  0.47 &  0.59 &  0.28 &  -- \\
      NGC~7790 &  3781 &                      CE~Cas~A &   1.09 &   0.50 &         $\varpi$, $\mu_{\alpha}^{*}$, $\mu_{\delta}$ &  1.00 &  0.28 &  0.28 &  -- \\
   Ruprecht~79 &  1701 &                        CS~Vel &   0.91 &   0.84 &  $\varpi$, $V_r$, $\mu_{\alpha}^{*}$, $\mu_{\delta}$ &  1.00 &  0.23 &  0.23 &  -- \\
     Loden~143 &  1807 &              OGLE-GD-CEP-0507 &  26.32 &   2.74 &         $\varpi$, $\mu_{\alpha}^{*}$, $\mu_{\delta}$ &  0.21 &  0.95 &  0.20 &  -- \\
       UBC~290 &    -- &                         X~Cru &   0.95 &   2.04 &         $\varpi$, $\mu_{\alpha}^{*}$, $\mu_{\delta}$ &  0.45 &  0.41 &  0.18 &  -- \\
   Gulliver~29 &    -- &              OGLE-BLG-CEP-172 &   0.97 &   2.20 &         $\varpi$, $\mu_{\alpha}^{*}$, $\mu_{\delta}$ &  0.39 &  0.47 &  0.18 &  -- \\
         BH~99 &  1831 &              OGLE-GD-CEP-0507 &  26.32 &   2.99 &         $\varpi$, $\mu_{\alpha}^{*}$, $\mu_{\delta}$ &  0.21 &  0.85 &  0.18 &  -- \\
       UBC~406 &    -- &                        CG~Cas &   1.03 &   1.72 &         $\varpi$, $\mu_{\alpha}^{*}$, $\mu_{\delta}$ &  0.58 &  0.30 &  0.17 &  -- \\
   Teutsch~145 &  2978 &           GDS~J1842359-051557 &   1.00 &   1.66 &         $\varpi$, $\mu_{\alpha}^{*}$, $\mu_{\delta}$ &  0.59 &  0.29 &  0.17 &  -- \\
        LP~699 &    -- &                        DK~Vel &   0.99 &   3.29 &         $\varpi$, $\mu_{\alpha}^{*}$, $\mu_{\delta}$ &  0.17 &  0.97 &  0.17 &  -- \\
       UBC~553 &    -- &              OGLE-GD-CEP-1196 &   1.01 &   0.90 &         $\varpi$, $\mu_{\alpha}^{*}$, $\mu_{\delta}$ &  1.00 &  0.16 &  0.16 &  -- \\
        UBC~80 &    -- &           ASAS~J060722+0834.0 &  15.53 &   3.24 &         $\varpi$, $\mu_{\alpha}^{*}$, $\mu_{\delta}$ &  0.18 &  0.91 &  0.16 &  -- \\
    Schuster~1 &  1756 &           GDS~J1004164-555031 &   1.07 &   2.30 &         $\varpi$, $\mu_{\alpha}^{*}$, $\mu_{\delta}$ &  0.36 &  0.44 &  0.16 &  -- \\
       LP~1332 &    -- &                        VV~Cas &   2.14 &   2.63 &         $\varpi$, $\mu_{\alpha}^{*}$, $\mu_{\delta}$ &  0.29 &  0.55 &  0.16 &  -- \\
       UFMG~69 &    -- &              OGLE-BLG-CEP-057 &   1.11 &   2.98 &         $\varpi$, $\mu_{\alpha}^{*}$, $\mu_{\delta}$ &  0.21 &  0.62 &  0.13 &  -- \\
     Loden~153 &  1824 &                        CS~Car &   1.08 &   1.71 &         $\varpi$, $\mu_{\alpha}^{*}$, $\mu_{\delta}$ &  0.46 &  0.28 &  0.13 &  -- \\
       UFMG~70 &    -- &              OGLE-BLG-CEP-057 &   1.11 &   1.91 &         $\varpi$, $\mu_{\alpha}^{*}$, $\mu_{\delta}$ &  0.49 &  0.27 &  0.13 &  -- \\
      NGC~4609 &  2062 &         WISE~J124231.0-625132 &   1.42 &   1.01 &         $\varpi$, $\mu_{\alpha}^{*}$, $\mu_{\delta}$ &  0.99 &  0.13 &  0.13 &  -- \\
 Collinder~228 &  1845 &              OGLE-GD-CEP-1673 &   1.03 &   3.66 &         $\varpi$, $\mu_{\alpha}^{*}$, $\mu_{\delta}$ &  0.13 &  0.93 &  0.12 &  -- \\
       UBC~345 &    -- &                     V0459~Sct &   0.88 &   3.77 &         $\varpi$, $\mu_{\alpha}^{*}$, $\mu_{\delta}$ &  0.12 &  1.00 &  0.12 &  -- \\
        LP~699 &    -- &           GDS~J0909005-533555 &   1.11 &   2.62 &         $\varpi$, $\mu_{\alpha}^{*}$, $\mu_{\delta}$ &  0.29 &  0.41 &  0.12 &  -- \\
       UBC~409 &    -- &                     V0824~Cas &   1.11 &   2.71 &         $\varpi$, $\mu_{\alpha}^{*}$, $\mu_{\delta}$ &  0.27 &  0.39 &  0.10 &  -- \\
     Loden~143 &  1807 &              OGLE-GD-CEP-1669 &  12.61 &   3.28 &         $\varpi$, $\mu_{\alpha}^{*}$, $\mu_{\delta}$ &  0.13 &  0.82 &  0.10 &  -- \\
       UBC~286 &    -- &              OGLE-GD-CEP-1707 &   1.16 &   1.36 &         $\varpi$, $\mu_{\alpha}^{*}$, $\mu_{\delta}$ &  0.76 &  0.14 &  0.10 &  -- \\
        LP~925 &    -- &  ASASSN-V~J062542.07+082944.4 &   1.16 &   2.95 &         $\varpi$, $\mu_{\alpha}^{*}$, $\mu_{\delta}$ &  0.22 &  0.45 &  0.10 &  -- \\

\noalign{\smallskip}
\hline

\end{tabular}
}
\vspace{0.001cm}
     {\raggedright 
     \\
     {\scriptsize $^*$ Uncertain Cepheid classification, as noted by the OGLE team. }\\
     {\scriptsize Open clusters with identification names starting with UBC correspond to clusters found by \citet{Castro-Ginard18,Castro-Ginard19,Castro-Ginard20}. The cluster names starting with LP are discoveries of \citet{Liu19}, whereas those starting with UFMG are from \citet{Ferreira21}. }
     {\scriptsize Cepheid names are taken from the International Variable Star Index \citep[VSX,][]{Watson06,Watson14}, or from the OGLE catalogue \citep{Udalski18}. 
     }
      \par}
\end{table*}

\section{Membership determination: The outcome}
\label{sec:outcome}

\par Of the total $\sim$44,300 possible combos for which we computed the likelihood of membership, only a small fraction displays relatively high probabilities, and the sample is strongly dominated by cluster-Cepheid pairs with probabilities $\leq 10^{-5}$. From the cluster-Cepheid pairs with higher membership probabilities, 163 have probabilities higher than 1\,per cent, 67 have probabilities over 10\,per cent, and only 44 over 25\,per cent. The thresholds mentioned here are only meant to give an overview of the results and bear no implication on membership, as our methodology relies on a hypothesis test that assumes membership in the first place. Cluster-Cepheid pairs with posterior probabilities higher than 0.10 are shown in Table~\ref{tab:combos}. Those with membership probabilities from 0.01 to 0.10 can be found in the appendix (Table~\ref{tab:combosA2}).

In Section~\ref{sec:bonafide} we briefly discuss a few combos from the literature that we consider recovered, relying mostly on the cluster Cepheid catalogues from David Turner\footnote{http://www.ap.smu.ca/$\sim$turner/cdlist.html}, A13, and \citet{Chen15}. In Section~\ref{sec:missing}, we briefly discuss six combos reported in at least three previous studies (from the aforementioned ones plus \citealt{Roeck12}) for which we obtain marginal membership probabilities. In Table~\ref{tab:combosA1} we show the results of these comparisons with literature combos. In Section~\ref{sec:new} we discuss a few arbitrarily selected combos.

\subsection{A few bona fide combos from the literature}
\label{sec:bonafide}

\subsubsection{The Cepheids around NGC~7790}

\par We recover the three Cepheids CE~Cas~A, and CE~Cas~B, CF~Cas paired to the cluster NGC~7790 (widely known as the only Galactic open cluster hosting three Cepheids), with relatively high association probabilities. The three Cepheids are bright (G $\sim$ 10\,mag) fundamental-mode pulsators with similar periods of $\sim$5\,d \citep{Ripepi19}, indicating they have a similar age. 

\par Our algorithm also reports a non-negligible probability of association of these three Cepheids with LP~888 \citep[][even with a slightly larger probability for CE~Cas~A]{Liu19} or with UBC~404. 
Although they are close to each other in the vicinity of NGC~7790, LP~888 and UBC~404 are reported as different structures by  \citet{Liu19} and \citet{Castro-Ginard20}, respectively. 
From previous knowledge \citep[e.g.,][]{Sandage58,Mateo88,Matthews95,Majaess13b} and given $P(A)$=1, the three Cepheids are clearly members of NGC~7790, but the quite high likelihoods computed suggest a dynamical association between NGC~7790 and the other structures newly discovered nearby. 

\subsubsection{The Cepheids in NGC~6067}

\par A second case of well-known cluster-Cepheid associations are the Cepheids V0340~Nor and QZ~Nor and the cluster NGC~6067. An extensive discussion addressing the possible membership of, especially, QZ~Nor is available in the literature \citep[see e.g., ][]{Eggen80,Walker85b,Coulson85,An07,Turner10,Majaess08}. 
A dedicated study performed by \citet{Majaess13a} confirmed both stars as members of NGC~6067. Recently, \citet{Breuval20} interpreted the proper motion difference between the cluster and QZ~Nor as a hint that the Cepheid is leaving the cluster. 
The striking difference in the membership probability (66\,per cent for V0340~Nor versus $<$ 1\,per cent for QZ~Nor) is a strong indication that the dynamical state of the cluster can have a strong impact on the membership probability. This is the reason why we provide a list of potential combos with a low membership probability, ranging from 1 to 10\,per cent (Table~\ref{tab:combosA2}) as it may contain similar cases. Finally, \citet{Breuval20} proposes GU~Nor as a potential member of NGC~6067 as well. We find a posterior probability $<0.01$ for this pair, as both its prior and likelihood are not significant. 

\subsubsection{GQ~Vul and FSR 0158}

\par For the combo consisting of the distant open cluster FSR~0158 \citep{Froebrich07} and the Cepheid GQ~Vul we also obtain a high association probability (0.47). 
This result is a combination of the position of the Cepheid close to the cluster's centre and the excellent agreement between their proper motions and parallaxes. It had been reported so far only by A13 with a probability of 43\,per cent.
We note, however, a discrepancy between the distance of FSR~0158 according to CG20 ($\sim6,100$\,pc) and the distance of GQ~Vul ($\sim4,500$\,pc) derived by \citet{Wang18} using a PL relation in the mid-infrared. 
Similarly, GQ~Vul is about $\sim$ 35\,Myr old from the theoretical PA relation of \citet{Bono05}, whereas the cluster age as determined by CG20 is $<$ 10\,Myr, a value incompatible with the presence of a Cepheid. However, it is noteworthy that only 27 stars are considered as cluster members with a probability higher than 50\,per cent, and only 14 with probabilities higher than 70\,per cent (CG18b, CG20), which may significantly affect the determination of FSR~0158's distance and age.\\

\par Combos for which we obtain $P(A|B)$~$>$ 0.01 and that appear at least once in Turner's database, A13 (with membership probability $>$ 0.10 from their work), or \citet{Chen15}, together with the spectroscopically confirmed cluster Cepheids described by \citet{Lohr18} and \citet{Clark15}, and the association recently found by \citet{Negueruela20}, are listed at the top of Table~\ref{tab:combosA1} (19 in total). 

\subsection{Missed combos from the literature}
\label{sec:missing}

\par Beyond the fact that combos previously reported in the literature might simply be discarded in the light of new, more accurate data, we could be unable to recover real combos for several reasons:
\par We exclude the fact that a cluster, and especially a Cepheid are missing in our catalog. Both the lists of clusters and Cepheids have been regularly updated and are much larger than the ones used for previous studies. Of course, there is always the possibility that an object is retracted, and this is actually the case for the Cepheid ASAS~J155149$-$5621.8 \citep{Pojmanski04} located 0.1\,deg away from the centre of the cluster NGC~5999, within its limiting radius ($r_{\rm lim} \sim 0.15$\,deg; \citealt{Ferreira19}). A potential association has been hypothesised by \citet{Chen15}, based on good agreement in proper motion, although they noted a mismatch for the computed age and distance modulus. However, the star is not considered as a classical Cepheid anymore: it is listed as a type~II Cepheid by \citet{Clementini19} and as "other" by \citet{Ripepi19}. It is considered a non-periodic variable in ASAS-SN \citep{Jayasinghe19b}.\\

\par Another possibility is that an insignificant membership probability originates from a low prior $P(A)$. The projected distance between the Cepheid and the cluster centre has obviously not substantially changed, and since we opted for a looser prior, the only possibility for this to happen is that the cluster apparent size has been modified after its core radius, limiting radius, or both, were modified. This could be the case for the potential association between the Cepheid X~Cyg and the cluster Ruprecht~175. This pair has been considered a bona-fide association by other authors in the past \citep[][]{Turner98a,Chen15}, but the pair's projected separation is 0.37\,deg (28.6\,pc, assuming membership), which, given a tabulated value of $r_1$=0.05\,deg for Ruprecht~175 (3.8\,pc, K13) gives a prior probability of virtually zero. We note in addition that the difference in parallax and proper motions between the cluster and the Cepheid are about ten times higher than their respective uncertainties, which leads to a negligible association probability.  \\

\par As mentioned above, updated values of the input parameters and their uncertainties with respect to those used in previous studies may result in smaller posterior probabilities, which may even become negligible and inconsistent with membership. This could have happened for the Cepheid BB~Sgr, associated with the cluster Collinder~394 by many authors \citep{Tsarevsky66,Turner84,Usenko19}. It has a relatively high prior $P(A)$=0.63, but there is a noticeable difference in the parallax and proper motions in right ascension and declination of the cluster and the Cepheid, which are 0.20\,mas, 1.83\,mas\ yr$^{-1}$, and 0.85\,mas\ yr$^{-1}$, respectively. An analogous case occurs for the combo WZ~Sgr and Turner~2, where the difference in parallax and proper motions in right ascension and declination of the cluster and the Cepheid is 0.22\,mas, 0.25\,mas\ yr$^{-1}$, and 0.57\,mas\ yr$^{-1}$ (respectively), the latter being larger than three times the cluster's proper motion dispersion. 
Similarly, the Cepheid CG~Cas has been considered for a long time a likely member of the cluster Berkeley~58 ($r_{50}=3.6$\,pc, \citealt{Turner08,Chen15}), of which it is separated by 0.09\,deg (5.4\,pc). In spite of having $P(A)=0.62$, we found a negligible membership probability for this pair due to their differences in parallax and proper motion.
Interestingly, we found instead higher probabilities for CG~Cas to be associated with UBC~406 (0.17) or LP~888 (0.09). 
In the case of CG~Cas and UBC~406, the angular separation corresponds to $1.7 \cdot r_{1}$ (0.17\,deg; 10.6\,pc, assuming membership), whereas the Cepheid is located at $2.7 \cdot r_{1}$ from the centre of LP~888 (0.74\,deg; 37.2\,pc). We note in addition that the Cepheid V0997~Cas shows signs of an association with the cluster LP~888, albeit with a low probability of 2\,per cent.

\par The case of RU~Sct is more complicated, because the Cepheid has been associated with the cluster Trumpler~35 \citep{Turner80,Chen15} as well as with other hosts, like Dolidze~32 (A13). In fact, A13 computed a membership probability of 0.52 with Dolidze~32 and of only 15\,per cent with Trumpler~35. In our study, we obtain a prior of 0.35 and 0.04 for the association of RU~Sct with Dolidze~32 and Trumpler~35, respectively, and posterior probabilities smaller than 1\,per cent in both cases. These insignificant $P(A|B)$ are mostly due to the large difference in proper motions, which exceed the uncertainties by about one order of magnitude, and due to the large radial velocity difference of 19\,km s$^{-1}$ in the case of the pair Dolidze~32 -- RU~Sct, where the individual uncertainties used are 7.4 and 2\,km s$^{-1}$, respectively. A third possible host for RU~Sct could be the cluster Dolidze~34, for which $P(A)=$ 0.07  (RU~Sct lies at approximately four times the cluster's $r_1$). 
However, in that case we also obtain an insignificant membership probability, based on a radial velocity, parallax, and proper motion comparison. \\

\par Moreover, we could end up with a very low posterior probability because uncertainties have been underestimated. This could for instance be the reason why we do not recover the Cepheid SU~Cyg associated with the cluster Turner~9 (\citealt{Turner98b, Turner10}; A13).
SU~Cyg is located at the centre of the cluster, which returns $P(A)$=1.0 in our analysis. However, parallax and proper motion differences between the pair are much larger than their corresponding uncertainties (about one order of magnitude), and we obtain therefore a small association probability for Turner~9 and SU~Cyg.

\par \citet{Hanke20} advocate for an additional unknown systematic error on Gaia DR2 proper motions. Analyzing stars with a possible globular cluster origin, they find that, from their position in the color-magnitude diagram and their absolute proper motion deviation with respect to the globular cluster M~13's mean value (see their Figure~2), the bright stars in their sample are obvious members of M13. However, those stars would not qualify as cluster members when taking into account the relative proper motion deviation. This result is a consequence of their membership likelihoods, computed from proper motions only and based on Mahalanobis distances, becoming very small for such stars. Since those stars have $G$ magnitudes of the order of $G=14-15$~mag, the effect might be similar for the open cluster members considered here, and even stronger for the somewhat brighter Cepheids in our sample. This could in turn artificially lower the value of our posterior probabilities.\\

\par Finally, we already mentioned in Section~\ref{sec:prior} that we adopted a looser prior as compared with A13 to accommodate for possible primordial or tidal features surrounding open clusters. 
However, it might be necessary to relax in addition the conditions related to parallaxes, proper motions, and radial velocities, in order to properly account for the dynamical state of the clusters. 
It could indeed be that the larger uncertainties in the pre-Gaia era were masking this effect, which could not be omitted anymore in the light of Gaia's accuracy and precision. We note in passing that inflating the uncertainties in the astrometric data by a factor of two would also increase the total number of combos with $P(A|B)$~$>$ 0.01 by a factor of two (from $163$ to $328$ pairs).\\

\subsection{Combos with high likelihood but low prior} 
\label{sec:new}

\par In Table~\ref{tab:combos}, where we list combos with probabilities higher than 0.10, the large majority of stars have a high prior. There are also a few stars with a lower prior, compensated by a high likelihood. In other words, their properties match extremely well with those of their potential host cluster, and they end up with a low probability only due to their large projected distance to the cluster. This is for instance the case of DK~Vel in LP~699, OGLE~GD-CEP-0507 in Loden~143 (MWSC~1807), SV~Per in ASCC~12 (MWSC~427), and V0459~Sct in UBC~345. 

\par When inspecting stars with lower membership probabilities (0.01$<P(A|B)<$ 0.10, Table~\ref{tab:combosA2}), the number of these cases increases, including for instance OGLE~GD-CEP-0964 and OGLE~GD-CEP-0968 in UFMG~54, RW~Cam in UPK~300, and OGLE~GD-CEP-1167 in UBC~545. A handful of such stars are associated with 2--4 clusters, but with high likelihoods only with 1--2 hosts, such as AQ~Pup and LP~1428, and OGLE~GD-CEP-1669 with Loden143~ and UBC~259.

\par A high likelihood is obviously not a guarantee of membership, as it can in particular be driven by issues in the determination of the astrometric parameters and/or large uncertainties.
Nevertheless, we highlight these cases as interesting pairs to further investigate. We note that if we do not restrict ourselves to combos with membership probabilities $P(A|B)$~$>$ 0.01, we find 258 additional combos with likelihood $P(B|A)$~$>$ 0.85. They are listed in Table~\ref{tab:combosA3} in the Appendix. Within this sample, 66\,per cent of the stars lie within 35 $\cdot r_{1}$, a value after which the distribution of projected radial distance drops drastically.    

\subsection{Some combos of interest}
\label{sec:new}

\par In this section we select arbitrarily a small number of combos for a more detailed discussion, focusing mostly on newly discovered clusters as potential hosts (not necessarily those with the highest membership probabilities).

\subsubsection{Clusters potentially hosting several Cepheids}

\par In our sample of combos with membership probabilities $P(A|B)$~$>$ 0.01, we find clusters that appear to be associated with several stars. We list them here below for further investigation and provide a few comments for each of them.
With the current data at hand, in addition to the astrometric and kinematic constraints used in this work, we conclude that only a couple of them are robust detections.

\par The Cepheids AQ~Pup and V620~Pup present probabilities of association with LP~1429 \citep{Liu19} of 0.04 and 0.01, respectively, mostly due to their likelihoods (0.42 and 0.40).
Possible associations of these Cepheids with overdensities, or putative clusters in their surroundings have been suggested in the past \citep{Turner12}, including Ruprecht~43, Ruprecht~44, and Turner~12. We report negligible membership probabilities in these cases, even considering the relatively high prior of AQ~Pup and Turner~12 (0.25). 

\par Three Cepheids are listed with non-negligible membership probability in BH~131: OGLE~GD-CEP-0785 is the closest one (at 17\,pc from the cluster centre, assuming membership) and therefore also has the highest prior ($\sim0.43$), while OGLE~GD-CEP-0790 and OGLE~GD-CEP-0795 lie farther away, with priors of 0.12 and 0.01. Only the latter has a higher likelihood (0.82), thus the three Cepheids have an overall membership probability of only 0.01--0.02.
Other clusters in the neighborhood of BH~131 are BH~132 and UBC~521. Both the priors and the likelihoods are negligible for the association of OGLE~GD-CEP-0785, OGLE~GD-CEP-0790, and OGLE~GD-CEP-0795 with these clusters.

\par Collinder 228 might host four Cepheids, namely V720~Car, OGLE~GD-CEP-0575, OGLE~GD-CEP-1672, and OGLE~GD-CEP-1673. They have priors $>$ 0.10 but relatively low likelihoods (with the exception of OGLE~GD-CEP-1673) and therefore end up with membership probabilities ranging from 1 to 12\,per cent.

\par For the Cepheids VY~Per, UY~Per, and SZ~Cas we report a $P(A|B)$ of 0.37, 0.35, and 0.07 with LP~2134 \citep{Liu19}, respectively, as a combination of their high priors and likelihoods. These relatively high membership probabilities make LP~2134 a case of interest, for which we consider further studies might be required.  
Other clusters within our list with which these Cepheids could be associated with because of their on-sky proximity, are Czernik~8, FSR~0591, UBC~190, ASCC~8, and SAI~17. 
However, for all of them the resulting $P(A|B)$ are near zero, including the pair Czernik~8--UY~Per, which has been considered a real association in previous works \citep[e.g., ][]{Turner77,Chen15}. 
For this pair in particular, the reason for its low probability is their low $P(B|A)$, which is not compensated by its slightly higher, but still poor prior (0.02\,per cent).

\par Five Cepheids are seemingly related to LP~699 (GDS~J0909005-533555, DK~Vel, V0530~Vel, OGLE-GD-CEP-0341, EX~Vel), with combinations of priors and likelihoods leading to membership probabilities from 1 to 17\,per cent, with the larger value corresponding to DK~Vel.
We note that four of these Cepheids, GDS J0909005-533555, DK~Vel, V0530~Vel, and OGLE-GD-CEP-0341, were not included in the analysis of A13, whereas EX~Vel was paired with the cluster Teutsch~48, although with a null membership probability from that work. We confirm this result. The other four Cepheids are initially crossmatched with other clusters in the field in our study. However, the membership probabilities of these pairs are not significant overall.

\par Finally, six Cepheids are potentially associated with LP~925, namely VW~Mon, V480~Mon, V966~Mon, ASAS~J062855+1107.3, OGLE~GD-CEP-0040, ASASSN-V~J062542.07+082944.4, 
but they have either a low prior or a low likelihood. Their membership probabilities range form 2 to 10\,per cent, making their association with LP~925 rather unlikely.

\subsubsection{Gaia 5 and V0423 CMa}

\par The Cepheid V0423~CMa lies within the half-light radius of the recently discovered cluster Gaia~5 \citep[2\,pc;][]{Torrealba19}. In the cluster discovery publication, the authors discard a possible association between Gaia~5 and V0423~CMa, arguing that the distance modulus difference determined in their study ($\sim$ 1.7) make the pair likely unrelated.
We analyzed the pair based on parallaxes only, since this is the only information available for both the cluster and the Cepheid. Our method outputs a membership probability of 0.94. Translated into distances, parallax values give a distance difference of $\sim$120\,pc only between the Cepheid and the cluster, a small value when compared with the cluster distance \citep[6.8\,kpc;][]{Torrealba19}.

\subsubsection{Kronberger~84 and ASASSN-V~J213533.70+533049.3} 

For the first overtone Cepheid ASASSN-V~J213533.70+533049.3 ($P=3.2$\,d) we find a possible association with the cluster Kronberger~84 (MWSC 3532, K13). In this case, the Cepheid lies close to the centre of the cluster (at $0.30$\,pc), well within its $r_{50}$ (0.02\,deg, 1.34\,pc). This results in $P(A)=1$. 
The posterior membership probability of this pair is 0.62, as a combination of both its high prior and likelihood. 
As a list of members of this cluster is provided by CG20, we display in Figure~\ref{fig:multiplots} the astrometry and color-magnitude diagram of this pair to illustrate its compatibility.

\subsubsection{UBC~130 and SV~Vul}

\par SV~Vul falls in a region of the sky with numerous clusters and star-forming regions, including Vul~OB1 with which the Cepheid has been associated for a long time \citep{Turner84}.  We find a high probability of association ($\sim$ 0.52) between SV~Vul and the open cluster UBC~130 \citep{Castro-Ginard20}. The distance of SV~Vul to the centre of UBC~130 is 6\,pc (0.15\,deg) assuming membership, about 50\,per cent larger than the cluster's $r_{50}$. It turns out that UBC~130 is another designation for the cluster Alicante~13, for which the membership of SV~Vul has been recently demonstrated by \citet{Negueruela20}. 
The astrometric parameters of SV~Vul as compared with those of the members of UBC~130 (CG20) are depicted in Figure~\ref{fig:multiplots}. 

\subsubsection{UBC~229 and V0335~Pup}

\par The angular separation between the Cepheid and the cluster's centre (0.04\,deg) corresponds to 1.8\,pc assuming membership, which locates V0335~Pup within the $r_{50}$ of UBC~229 (3.6\,pc) and secures a prior of $P(A)$=1. The constraints analysed in this case are the pair's parallaxes and proper motions (Figure~\ref{fig:multiplots}). They lead to an association probability of 0.51. Moreover, the position of the Cepheid in the CMD of UBC~229 and its distance are both compatible with membership.

\subsubsection{LP~1937 and DF Cas}

\par DF~Cas is located at 0.24\,deg of the centre of the newly discovered cluster candidate LP~1937 \citep{Liu19}, within the cluster's $r_2$ (0.486\,deg; 25.2\,pc).
The value of $P(A)$ for this pair is 0.64, 
and the small $\varpi$, $\mu^*_\alpha$, and $\mu_\delta$ differences between the Cepheid and the cluster yield a likelihood $P(B|A)$ of 0.71, hence a posterior membership probability of 0.45. 
We note the presence of other clusters in the neighborhood, such as NGC~1027, for which we get a membership probability $< 0.01$ ($P(A)$=0.77, $P(B|A)\sim$ 0), in agreement with the results of A13.

\subsubsection{UBC~106 and CM~Sct}

\par We find a prior of 0.79 and a likelihood of 0.40 for CM~Sct, hence a membership probability to UBC~106 of 0.32, based on parallax and proper motion. CM~Sct is located outside of the cluster's $r_{50}$ ($\sim$4.9\,pc or 0.12\,deg), with a physical distance of 6.6\,pc assuming the cluster and the Cepheid equidistant. The position of the Cepheid in the cluster's colour-magnitude diagram is compatible with membership (Figure~\ref{fig:multiplots}). We highlight this combo as A13 report a high likelihood for the combos CM~Sct/Dolidze~32 and CM~Sct/Dolidze~33, but with small membership probabilities of $<$ 1 and 1.6\,per cent, respectively. Their probability is slightly higher with Teutsch~145 (2.8\,per cent) thanks to a higher prior (0.21), but with a reduced likelihood of 0.13. We checked that the aforementioned clusters are distinct from UBC~106.

\par Our results indicate another Cepheid potentially associated with UBC~106, Z~Sct, which is located $\sim$700\,pc away in heliocentric distance from CM~Sct (and half a degree away on the sky) and has similar likelihood but a much lower prior than CM~Sct. With an angular separation between Z~Sct and the centre of UBC~106 (0.48\,deg, which corresponds to 19.7\,pc), the membership probability drops below 0.01. We note that the period of Z~Sct is significantly larger than that of CM~Sct, making it significantly younger. Our determination of the age of UBC~106 shows a quite large uncertainty, but overall matches log($t$) $=8.2$ provided by CG20. It then also supports a higher membership probability for CM~Sct than for Z~Sct, whatever the PA relation we consider. We note in passing that A13 mention Z~Sct in eight potential combos, all of them with negligible membership probability although with a likelihood $P(B|A)= $ 1 for three of them  (Dolidze~32, Dolidze~33, Andrews-Lindsay~5).

\subsubsection{UBC~290 and X~Cru}

The combo composed of the cluster UBC~290 \citep{Castro-Ginard20} and the fundamental-mode pulsator X~Cru ($P = 6.22$\,d) is another case of a relatively high $P(A|B)$ association, as the Cepheid lies at ~8.6\,pc from the center of the cluster of size $r_{50} = 0.15$\,deg (4.24\,pc), which, together with a high likelihood, yields a posterior probability of 0.18 (Figure~\ref{fig:multiplots}). 
We note that neither the cluster nor the Cepheid are included in the study of A13.

\section{Age Determination of open clusters}
\label{sec:ages}

In this section, we derive ages for a subsample of clusters believed to host Cepheids in the literature, or where the Cepheid has a high membership probability according to our study (see Section~\ref{sec:outcome}, and Table~\ref{tab:combos}). We compare these estimates with age determinations from the literature, and check their consistency with theoretical Cepheid pulsation and evolution models. 

\subsection{Methodology}
\label{sec:methodology}

\par Age-dating resolved star clusters via isochrone fitting is a task for which several techniques have been used along the years, from pure visual inspection to recently developed algorithms
(see e.g., \citealt{vonHippel06,Monteiro10,Dias12,Liu19,Sim19}; CG20)
In most cases, the codes used for these calculations are not made publicly available. 
However, in spite of the application of new methodologies, determining the age of young clusters remains as a challenging goal because in addition to stellar contamination, binarity, and age spreads, the MSTO of these clusters is commonly not clearly defined. 

\par We adopted two approaches: a $\chi^2$-based isochrone selection developed on our own, and the AURIGA neural network \citep[henceforth ANN;][]{Kounkel20}, which predicts the age, extinction, and distance of clusters from the photometry and astrometry of the cluster members. We did not use the software BASE 9 \citep{vonHippel06} as in \citet{Bossini19} since these authors mention that its use together with only Gaia magnitudes does not allow one to lift the degeneracy between the distance modulus and the extinction.\newline

\par For our own method, we used the PAdova and TRieste Stellar Evolution Code  \citep[PARSEC; ][]{Bressan12} stellar evolution models. The models were computed for the Gaia DR2 passbands \citep{Evans18}, and in the 2MASS photometric system.
From the available models, we selected evolutionary tracks with initial chemical compositions ranging from Z $ = 0.006$ to Z $ = 0.029$ and a grid size of $0.001$\,dex. 
Approximately 90\,per cent of the clusters in the catalogue of \citet{Carrera19} younger than 800\,Myr (with ages from CG20) lie in this range of Z.  
The ages selected vary from log($t$) $ = 6.6$ ($\sim 5$\,Myr), with $t$ in units of years, to log($t$) $ = 8.9$ ($\sim 800$\,Myr) with a minimum resolution of $0.01$\,dex, including extreme values of log($t$) for a proper uncertainty determination.  
Since rotationally-induced instabilities strongly affect the evolution of stars, as studied specifically in the case of Cepheids by \citet{Anderson16}, we also adopted models that take stellar rotation into account, namely the MESA Isochrones and Stellar Tracks (MIST) set of evolutionary tracks, which are based on the publicly available stellar evolution tool Modules for Experiments in Stellar Astrophysics \citep[MESA;][]{Paxton11,Paxton13,Dotter16,Choi16}. We chose evolutionary tracks with a range of initial iron abundances [Fe/H] from $-0.5$ to $0.5$ with a grid size of $0.25$, and logarithmic ages between $6.6$ and $8.9$, with a step size of $0.05$\,dex.

\par The selection of the best model isochrone for a given cluster is based on a $\chi^2$ minimization criterion when comparing it with the cluster colour-magnitude diagram \footnote{The Cepheids were excluded from the computation of the $\chi^2$ values.}. We repeated the process twice, first using Gaia photometry only and then using 2MASS photometry only. In the latter case, the near-infrared $J$, $H$, and $K$ magnitudes of the cluster members are taken from \citet{Roeser10}.
\par In order to perform the isochrone fitting, we limited ourselves to cluster members as established in previous studies, and we discarded clusters for which the main sequence was not clearly defined. In the case of the clusters in CG18b, who provide a membership probability, we included all stars with membership probabilities larger than 30\,per cent and within $2.5\cdot r_1$ from the cluster centre. For the clusters from \citet{Castro-Ginard20}, we simply used the list of members provided by the authors. Additionally, a number of stars was further excluded during the fitting process via sigma-clipping. 
As initial conditions for the fitting routine, we adopted the values provided in the literature. From this original value, we explored an age window of $\pm \ $0.5 in logarithmic scale, using $0.05$ as a grid size
and a metallicity window of $\pm \ 0.03$\,dex in Z, with a grid size of $0.01$\,dex. For the reddening, we allowed for an excursion of $\pm \ $ 0.6 mag from the initial value, in steps of $0.15$ mag. The extinction values were computed assuming $R_V = 3.1$ \citep{Schultz75,Cardelli89}, adopting the ratios $A_G/A_V = 0.85926$, $A_{G_{BP}}/A_V = 1.06794$, $A_{G_{RP}}/A_V = 0.65199$, $A_J/A_V = 0.29434$, $A_HG/A_V = 0.18128$, and $A_K/A_V = 0.11838$\footnote{http://stev.oapd.inaf.it/cgi-bin/cmd\_3.3} , and without considering differential reddening. Finally, we allow distances to vary in a range of $\pm \ $ 500\,pc from the initial value, in steps of 50\,pc. 
\par To account for the magnitude errors on the stars' $G$, $G_{\rm BP}$, and $G_{\rm RP}$ magnitudes, we assumed: 

\begin{equation}
\sigma_{\rm mag}^2 = (1.09 \ \frac{\sigma_{Flux}}{Flux})^2 + \sigma_{\rm zp}^2
\end{equation}

\noindent where $\sigma_{\rm mag}$ is the magnitude error of a star in a given Gaia bandpass, $Flux$ is the mean flux of the star in that filter, and $\sigma_{Flux}$ its uncertainty. We used this formula since no errors are provided in the Gaia catalogues because of the asymmetric error distribution of the sources in magnitude space, as stated in the table description of the Gaia DR2\footnote{https://dc.zah.uni-heidelberg.de/tableinfo/gaia.dr2light}. 
We adopted $\sigma_{\rm zp} = 0$ as we possess no knowledge of the behaviour of this zero-point, and its effects should be negligible for the purpose of our study. 
\par The uncertainties on the age determination were estimated by inspecting the distribution of the minimum $\chi^2$ as a function of age in the range explored for a given cluster. We computed the significance of the global minimum by looking for the local maxima around it, allowing asymmetric errors if necessary. 

\par
\begin{figure}
\begin{center}
\includegraphics[angle=0,scale=.30]{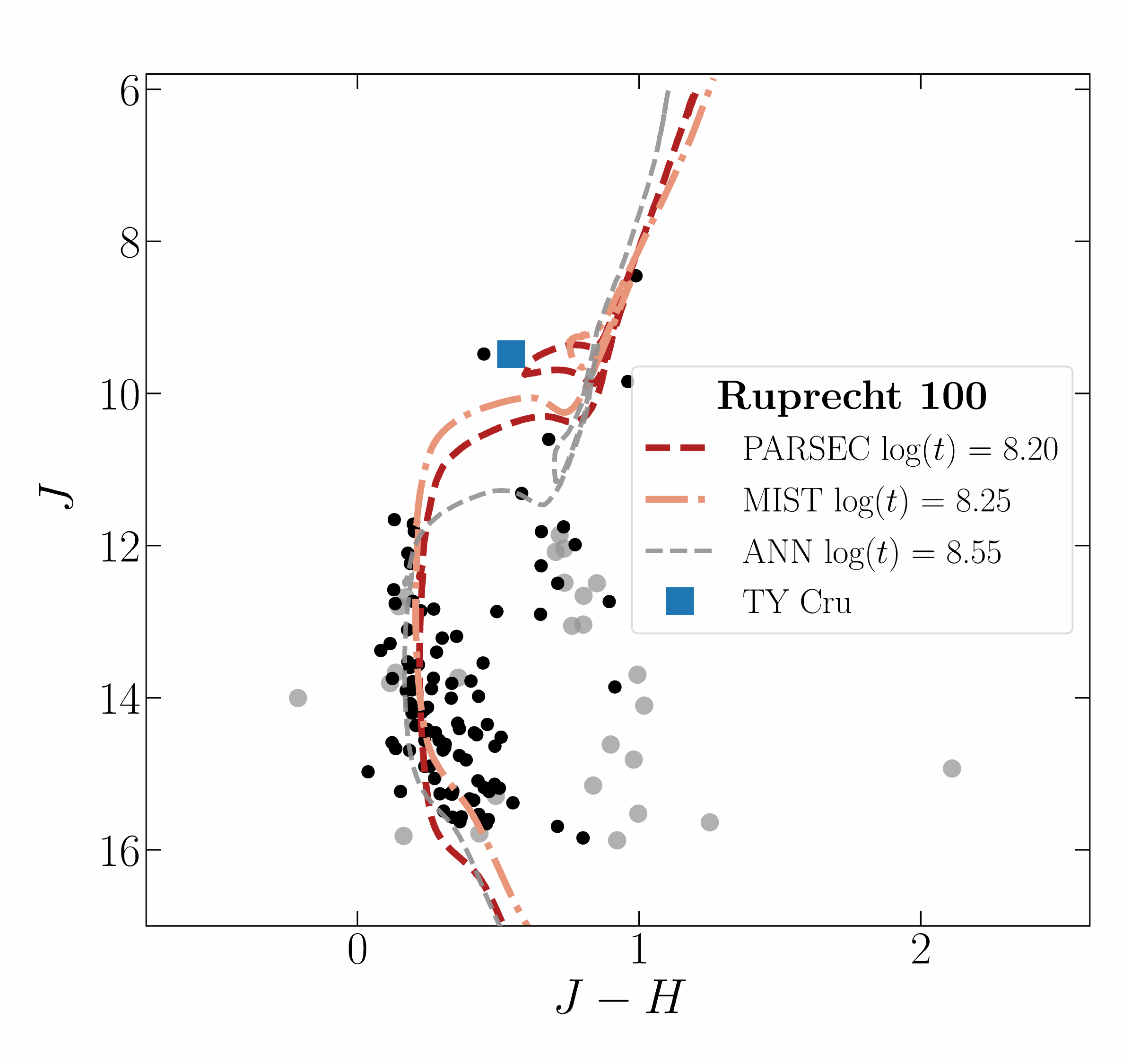}
\caption{Color-magnitude diagram of the cluster Ruprecht~100, with PARSEC (red) and MIST (orange) isochrones fitted (as an exception) by visual inspection.
Cluster members are taken from CG18b and shown as grey filled circles, while the cleaned sample of stars used during the isochrone fitting procedure is shown as black dots. For this cluster, automatic fits based on a $\chi^2$ minimization did not converge to an acceptable solution, even when allowing for a manual shift of the age. A PARSEC isochrone computed with the parameters derived by the ANN analysis and assuming solar metallicity is also plotted in grey. }
\label{fig:CMDsByHand}
\end{center}
\end{figure}

 In a few cases, when the fit appeared inconsistent with the cluster members in a visual inspection, we applied small adjustments, setting the age to a local rather than a global minimum or correcting for small distance/reddening imprecisions when the best value would fall in between two consecutive grid points. We did so for seven clusters analyzed with Gaia photometry and PARSEC isochrones, five clusters with Gaia photometry and MIST isochrones, two clusters with 2MASS photometry and PARSEC isochrones, and four clusters with 2MASS photometry and MIST isochrones. The median shift in age is 0.20 in logarithmic scale, with a maximum of 0.60 for the cluster Ruprecht~100 in a Gaia+PARSEC configuration. The case of Ruprecht~100 in the 2MASS+PARSEC configuration remained nevertheless an unsatisfactory fit and was fitted using a visual inspection only, as displayed in Figure~\ref{fig:CMDsByHand}.\newline

Alternatively, we employed the  ANN\footnote{https://github.com/mkounkel/Auriga} to derive the cluster properties. The ANN is a neural network trained on a mix of artificial stellar populations and real clusters. We provided as input for each cluster photometry in the Gaia and near-infrared bands ($G$, $BP$, $RP$, $J$, $H$, $K$), and the Gaia DR2 parallaxes. 
\citet{Kounkel20} indicate that the ANN underestimates the age for clusters older than $\sim$120~Myr and overestimates it for clusters younger than $\sim$120~Myr, in both case by $\sim$0.1 dex. They remark that this threshold roughly corresponds to the age at which all low-mass pre-main sequence stars would have reached the main sequence, and very few high-mass stars would have evolved off the mains sequence towards the RGB. Unfortunately, this is also the expected age range for clusters hosting Cepheids. In general, neural networks do not provide uncertainties in the predicted parameters. However, it is possible to treat the scatter between the solutions from independent realizations as a measure of these errors. Thus, for each cluster we ran 100 ANN iterations to estimate the parameter uncertainties. For more details about the ANN design, we refer the reader to \citet{Kounkel20}.

\begin{figure*}
\begin{center}
\includegraphics[angle=0,scale=.30]{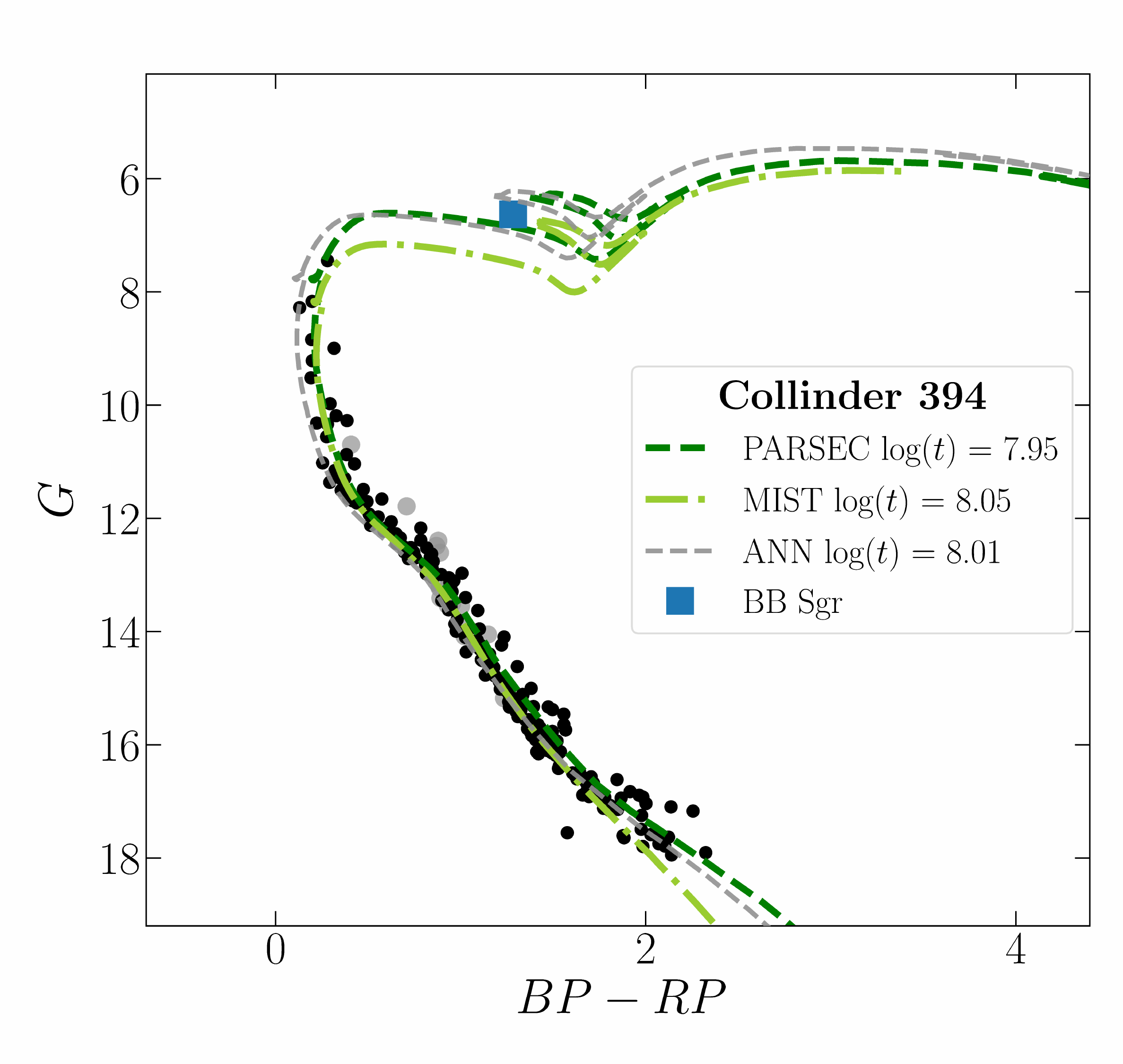}
\includegraphics[angle=0,scale=.30]{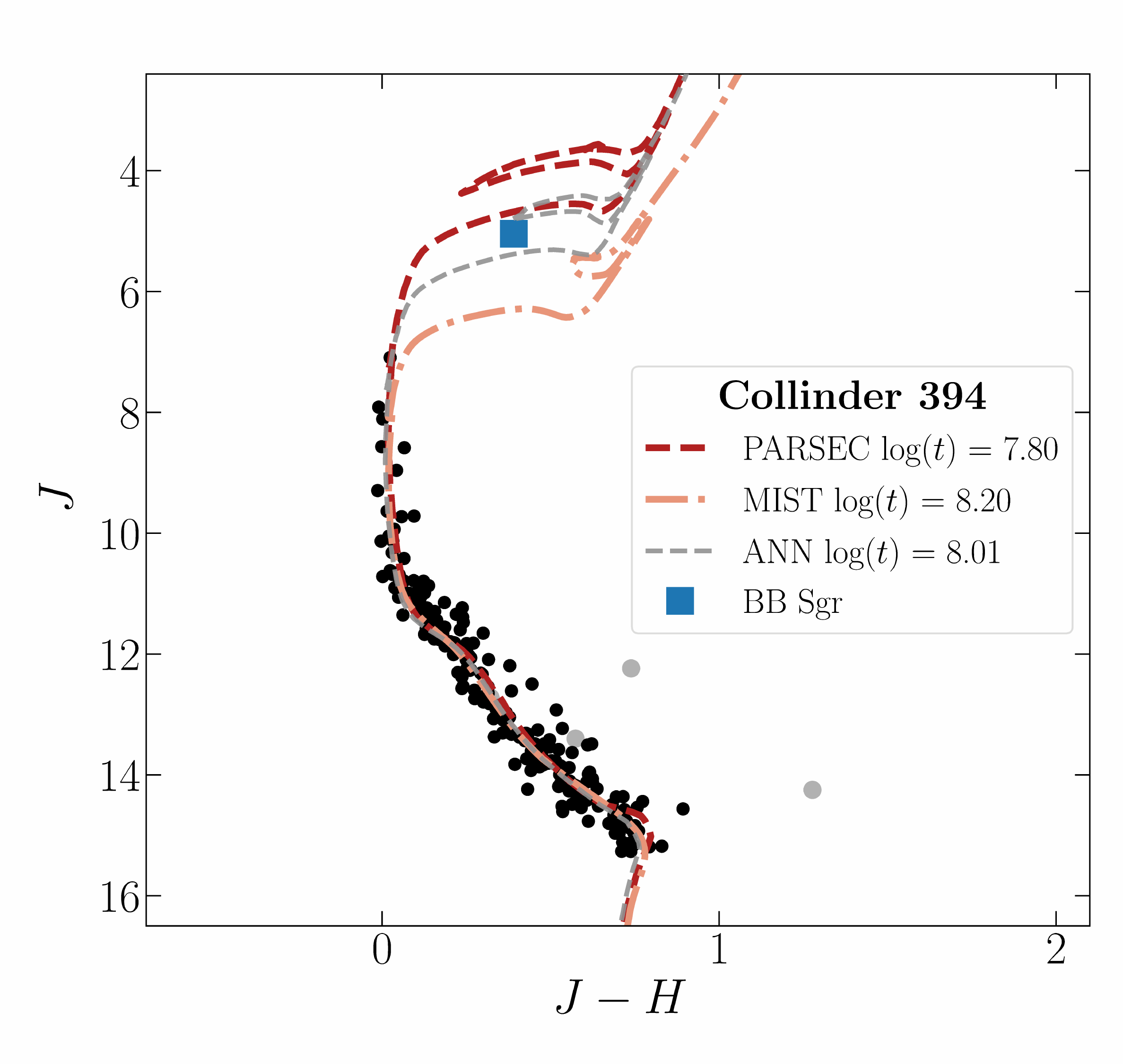}
\includegraphics[angle=0,scale=.30]{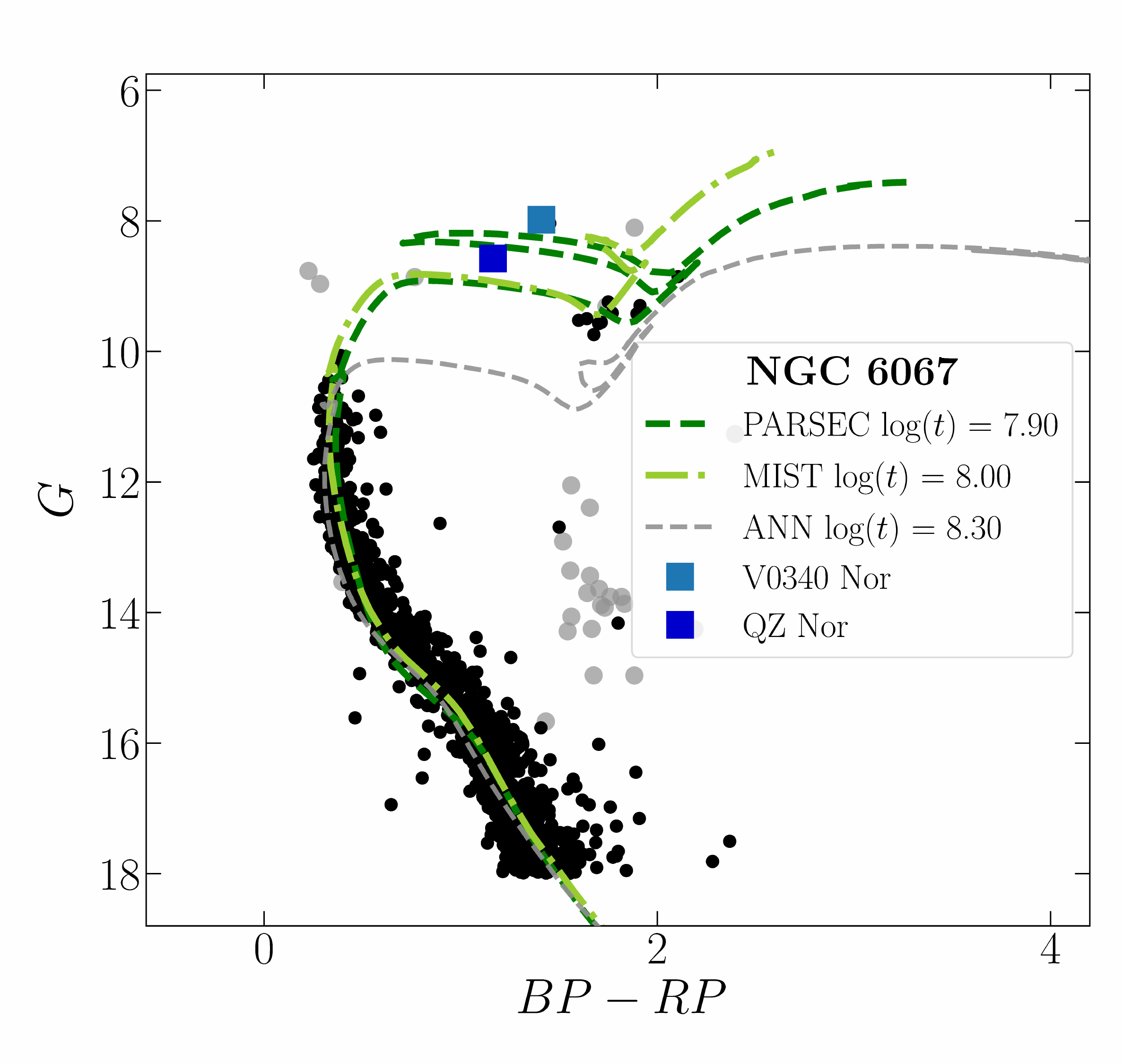}
\includegraphics[angle=0,scale=.30]{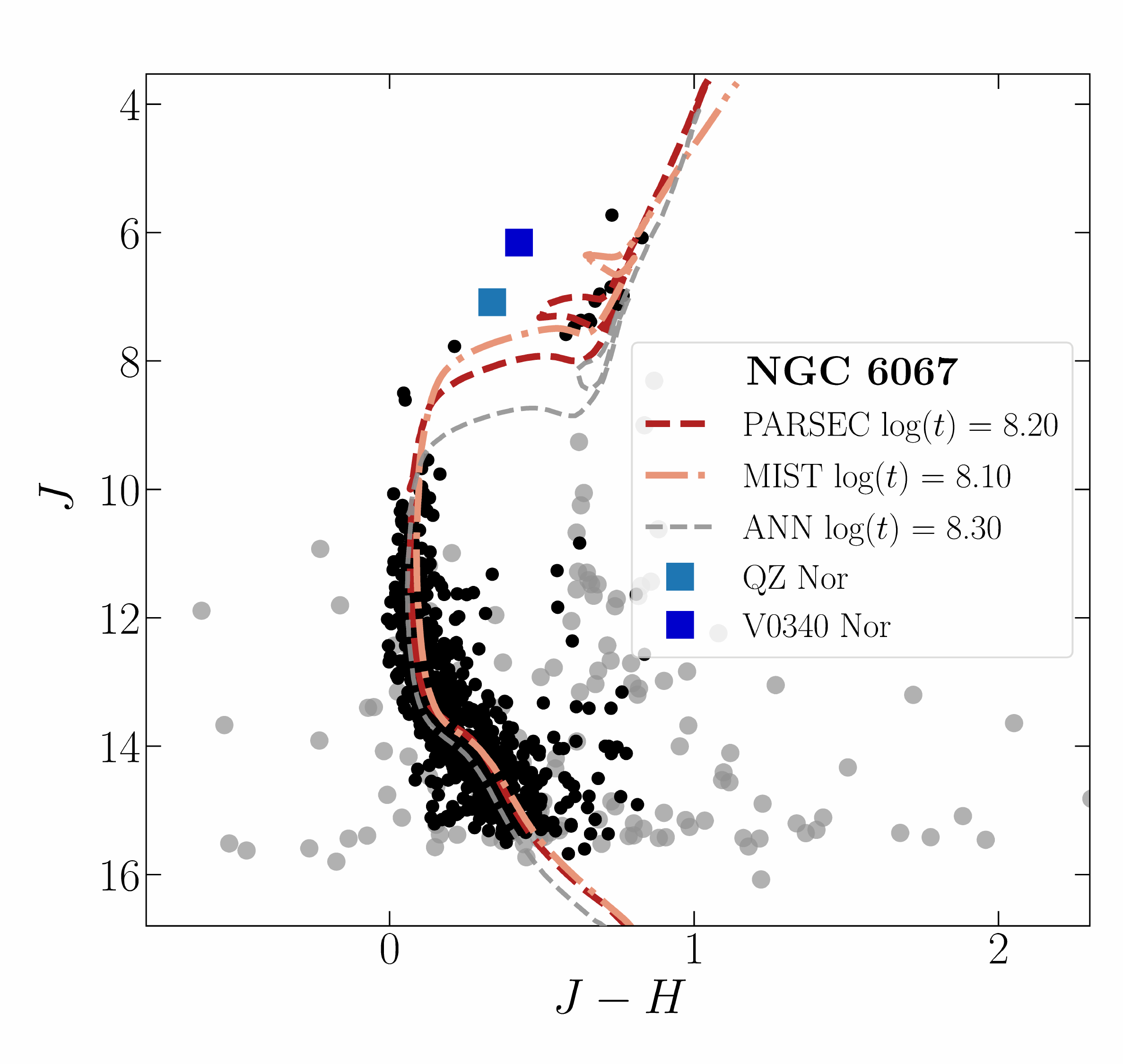}

\caption{
Colour-magnitude diagrams of two clusters: Collinder 394 (top panels) and NGC 6067 (bottom panels) representative of clusters hosting Cepheids. The left panels show colour-magnitude diagrams in the Gaia passbands while the right panels display them in the 2MASS passbands. Presumed cluster members from the input catalogues are shown as grey filled circles while those selected by sigma-clipping during the isochrone fitting procedure are shown as black dots.  The best-fit isochrones are shown with a colour-code related to the photometric system and the stellar evolution model used in the analysis. We also show in grey a PARSEC isochrone computed with the parameters derived by the ANN analysis and assuming solar metallicity. }
\label{fig:CMDs}
\end{center}
\end{figure*}

\subsection{Results and comparison with previous studies}
\label{sec:results}

\par It is no surprise that our analysis of clusters hosting Cepheids provides only young ages, ranging from log($t$)=7.4 (25~Myr) to log($t$)=8.8 (630~Myr), depending on the cluster, the method and the data considered. For reasons that will become clear later in this Section, we discuss here and list in Table~\ref{tab:ages} only 11 clusters (12 Cepheids) whose ages are considered relatively reliable (considering the accuracy and precision of the estimations) and with which we could potentially constrain the Cepheid period-age relation (see Section~\ref{sec:PAs}). Figure~\ref{fig:CMDs} displays the results for two representative clusters. The first one, NGC~6067, is a relatively evolved system with a well-defined main sequence, two Cepheids, and already a number of stars populating the RGB. With such favorable circumstances, all analyses lead to a similar age between log($t$)=7.9 (80~Myr) and log($t$)=8.3 (200~Myr). 
The second one, Collinder~394, is in contrast less massive, it contains only one Cepheid and no other evolved member has been reported in Gaia DR2 so far. Fortunately, it possesses a narrow main sequence and a quite well defined MSTO, allowing for a similar dispersion ($\sim$ 0.4\,dex) around an age of log($t$)$\approx$8.05 (110~Myr). The resulting isochrones and cluster colour-magnitude diagrams of our whole sample are shown in the Appendix (Figure~\ref{fig:allCMDs1}). It is clear that our "good" sample is biased toward higher ages since we report log($t$)$<$ 7.8 for only two clusters.

\begin{table*}
\small
\caption{
Ages for a reliable subsample of clusters. 
We show the logarithmic cluster ages obtained with two sets of isochrones (PARSEC or MIST) using either Gaia DR2 or 2MASS photometry, and those obtained with ANN using Gaia DR2 and 2MASS simultaneously. 
}
\smallskip
\begin{center}
\label{tab:ages}
{\small
\begin{tabular}{ccccccccccc}
\hline
\noalign{\smallskip}
Open cluster & Cepheid & Period$^1$ & Pulsation &  \multicolumn{5}{c}{log($t$)} & \multicolumn{2}{c}{Lit. log($t$)} \\
 & &  & Mode$^1$ & GDR2 & GDR2 & 2MASS & 2MASS & \\
  & &   &  & PARSEC & MIST & PARSEC & MIST & ANN & CG20$^2$ & \citet{Bossini19} \\
 & & [days] &  &  &  &  &  & & \\
\noalign{\smallskip}
\hline
\noalign{\smallskip}
  Collinder 394 & BB Sgr & 6.64 & F & $7.95^{+0.20}_{-0.40}$ & $8.05^{+0.20}_{-0.25}$ & $7.80^{+0.30}_{-0.30}$ & $8.20^{+0.15}_{-0.15}$ & $8.01\pm{0.09}$ & 7.96 & $7.97^{+0.06}_{-0.02}$ \\
  NGC 5662 & V Cen & 5.49 & F & $7.95^{+0.20}_{-0.15}$ & $7.95^{+0.15}_{-0.15}$ & $7.85^{+0.25}_{-0.25}$ & $8.45^{+0.30}_{-0.20}$ & $8.22\pm{0.12}$ & 8.30 & --\\
  NGC 6067 & QZ Nor & 3.79 & 1O & $7.90^{+0.50}_{-0.30}$ & $8.00^{+0.35}_{-0.40}$ & $8.20^{+0.30}_{-0.35}$ & $8.10^{+0.45}_{-0.20}$ & $8.30\pm{0.14}$ & 8.10 & --\\
  NGC 6067 & V0340 Nor & 11.29 & F & $7.90^{+0.50}_{-0.30}$ & $8.00^{+0.35}_{-0.40}$ & $8.20^{+0.30}_{-0.35}$ & $8.10^{+0.45}_{-0.20}$ & $8.30\pm{0.14}$ & 8.10 & --\\
  NGC 6087 & S Nor & 9.75 & F & $7.75^{+0.70}_{-0.70}$ & $7.90^{+0.20}_{-0.20}$ & $7.90^{+0.35}_{-0.35}$ & $8.05^{+0.55}_{-0.55}$ & $7.92\pm{0.10}$ & 8.00 & $8.05^{+0.02}_{-0.03}$ \\ 
  NGC 6649 & V367 Sct & 6.29 (4.38) & F1O & $7.45^{+0.20}_{-0.20}$ & $7.40^{+0.60}_{-0.60}$ & $7.75^{+0.20}_{-0.20}$ & $7.70^{+0.35}_{-0.35}$ & {$7.18\pm{0.39}$}$^{\ding[0.6]{60}}$ & 7.85 & --\\ 
  Ruprecht 79  & CS Vel & 5.90 & F & $7.80^{+0.25}_{-0.20}$ & $7.80^{+0.35}_{-0.35}$ & $7.90^{+0.60}_{-0.60}$ & $7.60^{+0.55}_{-0.40}$ & $7.77\pm{0.24}$ & 7.79 & --\\
  Ruprecht 100 & TY Cru & 5.00 & F & $8.40^{+0.10}_{-0.60}$ &
  $8.00^{+0.55}_{-0.10}$ &$^{\ding[0.7]{61}}${$8.20^{+0.55}_{-0.55}$} & $8.25^{+0.25}_{-0.45}$ & $8.55\pm{0.23}$ & 8.31 & --\\ 
  UBC 106 & CM Sct & 3.92 & F & $8.20^{+0.30}_{-0.15}$ & $8.15^{+0.25}_{-0.15}$ & $8.00^{+0.35}_{-0.35}$ & $7.85^{+0.25}_{-0.25}$ & $7.96\pm{0.43}$ & 8.20 & --\\ 
  UBC 130 & SV Vul & 44.88 & F & $7.45^{+0.15}_{-0.40}$ & $7.20^{+0.15}_{-0.15}$ & $7.40^{+0.30}_{-0.30}$ & $7.20^{+0.40}_{-0.30}$ & {$7.53\pm{0.32}$}$^{\ding[0.6]{60}}$ & 7.44 & --\\
  UBC 156 & V1077 Cyg & 4.64 & F & $8.80^{+0.65}_{-0.65}$ & $8.45^{+0.30}_{-0.50}$ & $8.60^{+0.15}_{-0.55}$ & $8.80^{+0.35}_{-0.35}$ & $8.19\pm{0.18}$ & 8.40 & --\\
  UBC 290 & X Cru & 6.22 & F & $8.05^{+0.20}_{-0.35}$ & $7.90^{+0.35}_{-0.35}$ & $8.00^{+0.35}_{-0.35}$ & $7.80^{+0.10}_{-0.10}$ & $7.83\pm{0.11}$ & 8.28 & --\\
\noalign{\smallskip}
\hline
\end{tabular}
 \vspace{0.001cm}
 
     {\raggedright $^1$ From \citet{Ripepi19}, except for R~Cru, V367~Sct \citep[VSX,][]{Watson14} and for OGLE-GD-CEP-1012 \citep[OGLE,][]{Udalski18}.\\
     In the pulsation mode column, F stands for fundamental mode, 1O for first overtone, and F1O for double-mode. 
     For the double mode Cepheid V367~Sct the first overtone period is shown in parentheses.\par}
     
     {\raggedright $^2$ The uncertainties associated with the log($t$) determinations of CG20 range from $0.1$ to $0.25$\,dex. \par}
     
     {\raggedright $^{\ding[0.8]{61}}$ Unsatisfactory $\chi^2$ fitting procedure. Age determined by visual inspection. \par}
     
     {\raggedright $^{\ding[0.6]{60}}$ Clusters for which high values of the extinction $A_V$ ($>$1.6, CG20; $>$1.9, ANN) affected the ANN parameter determination. 
     
     \par} 
}
\end{center}
\end{table*}
\par
\begin{figure}
\begin{center}
\includegraphics[angle=0,scale=.375]{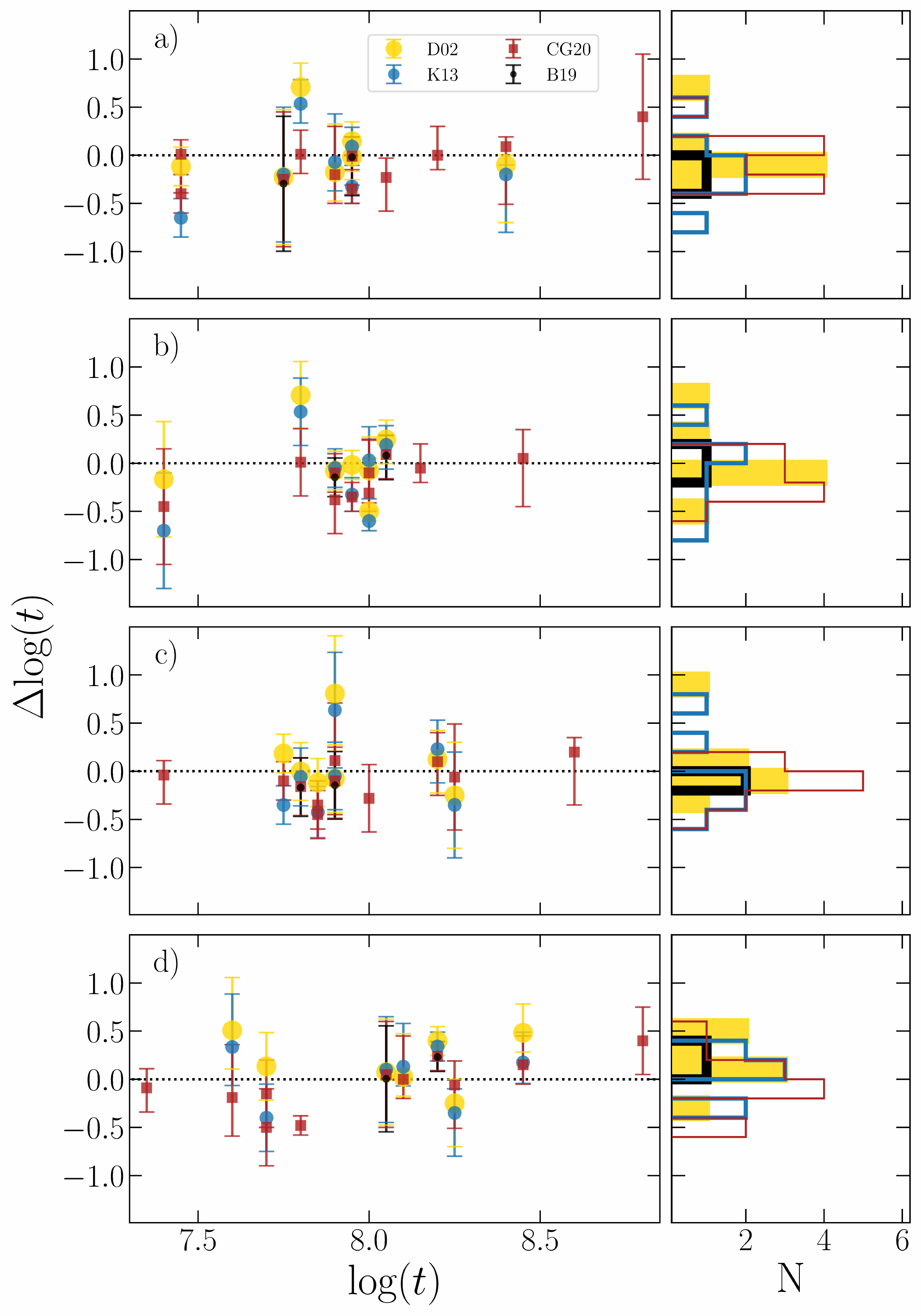}
\caption{{\it Left panels:} Difference in logarithmic ages, $\Delta$ log($t$), between the values obtained in this work and those published in previous studies, for the sub-sample of clusters shown in Table~\ref{tab:ages}. {\it Right panels:} Dispersion of the residuals in $\Delta$ log($t$) using a bin size of 0.25 in age logarithmic scale.
The sources of literature ages are D02 (yellow), K13 (blue), \citet{Bossini19} (B19, black), and CG20 (red). The panels from top to bottom display: 
\textit{a)}: our ages from PARSEC isochrones, using Gaia photometry,
\textit{b)}: our ages from MIST isochrones, using Gaia photometry,
\textit{c)}: our ages from PARSEC isochrones, using 2MASS photometry,
\textit{d)}: our ages from MIST isochrones, using 2MASS photometry. 
The error bars take into account the age uncertainties from this work and from the literature, when available.}
\label{fig:agesComp}
\end{center}
\end{figure}

 In the rest of this subsection we compare the age estimates from our first approach with previous studies, namely the compilations of D02 and K13, the analysis of \citet{Bossini19} who derived ages for 269 low reddening (not very young) Galactic open clusters using Gaia DR2 data (two clusters in common with our work), and the recent work of CG20 who used an artificial neural network to determine the ages of 1,867 clusters from Gaia DR2 photometry. The outcome of the comparison is shown in Figure~\ref{fig:agesComp}, from which we draw the following conclusions (which we emphasize are drawn from low number statistics):
\begin{itemize}
    \item we find an overall agreement between all the literature ages we compared our results with (within $0.8$\,dex);
    \item the agreement of our ages with D02 is better than that our agreement with K13, most likely due to the cluster membership selection. The median absolute differences are $0.16$ and $0.33$ in logarithmic scale, respectively, for the seven clusters in common with both works; 
    \item we reach an overall good agreement with the \citet{Bossini19} (two clusters in common) and CG20 age estimates (11 clusters in common, median absolute difference of $0.15$\,dex).
    However, for clusters younger than log($t$)$\sim$8.0 (100\,Myr), our ages tend to be lower than those of CG20, and higher for log($t$) $>$ 8.2 ($\sim$160\,Myr);  
    \item there are no other clear trends of $\Delta$ log($t$) as a function of log($t$);
    \item the choice of a specific isochrone set (PARSEC or MIST) seems to have marginal influence.
    
\end{itemize}
We believe that the age difference between earlier studies (as compiled by D02 and K13) and the more recent ones resides mainly in the capability to select the cluster membership based on Gaia DR2 parallaxes and proper motions. 
However, the assumed reddening and distance for each cluster (in both earlier and recent studies) likely play an important role as well. 
Finally, we consider important to keep in mind the typical values of $\Delta$ log($t$) when analysing the cluster ages from different works (and their spread), in particular for interpreting the results shown in Section~\ref{sec:PAs}.

\subsection{A critical view on cluster Cepheids to calibrate period-age relations}
\label{sec:critical}

\par
\begin{figure*}
\begin{center}
\includegraphics[angle=0,scale=.50]{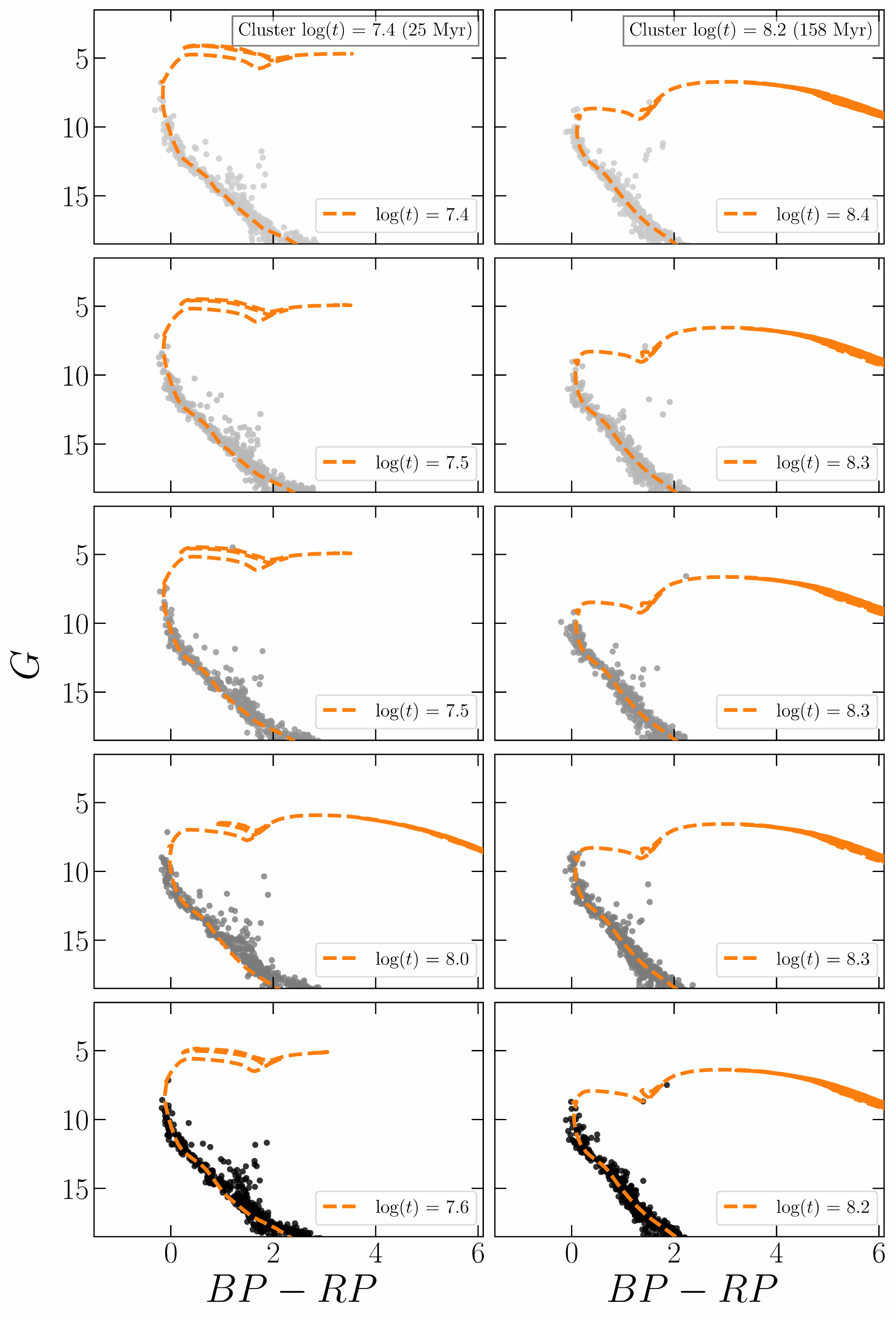} 
\caption{Simulated colour-magnitude diagrams for a cluster of log($t$) $=7.4$ (left panels) and log($t$) $=8.2$ (right panels), and a total initial mass of 1,000\,M$_\odot$. 
Each panel shows a different population randomly generated using a \citet{Chabrier01} + \citet{Salpeter55} IMF. Observational effects, such as photometric errors, the presence of a binary sequence, and field contamination are included in each diagram. PARSEC isochrones are fitted to the observed CMDs and displayed with orange dashed lines. The isochrones were fitted assuming the exact same distance, extinction, and metal content as those of the theoretical population, but allowing for shifts in age.}
\label{fig:TheoCMDs}
\end{center}
\end{figure*}

 To illustrate the difficulty of deriving accurate ages for young open clusters, we study the theoretical behaviour of the stellar occupation of the members of a given cluster in the Gaia colour-magnitude diagram, using simple models based on stellar population synthesis. We used PARSEC isochrones to generate ten simulated clusters of solar metallicity at two specific ages (five random realizations per age). The ages selected for these models are 7.4 to 8.2 in logarithmic scale (approximately 25 and 160\,Myr, respectively), for a massive Galactic open cluster (1,000\,$M_\odot$; based on the cluster mass functions shown by, e.g., \citealt{Lada03}, \citealt{Zinnecker09}, and \citealt{Roser10}). To compute the stellar occupation along the isochrones we adopted a \citet{Chabrier01} initial mass function (IMF), together with a \citet{Salpeter55} IMF for stars with masses larger than a solar mass. To transform the modeled absolute magnitudes to apparent magnitudes, we assumed a distance of 1\,kpc, and $E(B-V)= 0.175$, which are representative values for a Galactic cluster such as NGC\,6087. We included photometric uncertainties based on the typical magnitude errors for the NGC 6087 members in the Gaia passbands, re-drawing the magnitudes assuming Gaussian distributions. In addition, we adopted a binary fraction of 0.6 to roughly reproduce the characteristic widening produced in the evolutionary tracks by the presence of binary companions. Finally, we added field contamination (foreground/background) based on the Besan\c con models\footnote{https://model.obs-besancon.fr/modele\_home.php} \citep{Robin03,Czekaj14,Robin14} of Galactic stellar populations for the Gaia magnitudes, including only a random selection of stars with distances between 0.8 and 1.2\,kpc as possible contaminants. The results of this exercise are depicted in Figure~\ref{fig:TheoCMDs}.

\par Isochrones of different ages, in the ideal scenario in which the true distance, reddening, and metallicity of the cluster are known, are fitted to the populations plotted in the different panels of Figure~\ref{fig:TheoCMDs}. The scatter in the fitted logarithmic ages with respect to the isochrone ages from which the populations are drawn shows the sensitivity to different effects (e.g., photometric errors and contamination) of the age determination of a stellar cluster via isochrone fitting, and one expects it to only increase if small variations in the isochrone distances and reddenings were allowed. 

\par A factor that is not being considered here is related to the fraction of observed stars that are recovered from the theoretical stellar populations, due to the survey photometric completeness, or to possible biases in the cluster census made by the studies that selected cluster members. In the example populations depicted in Figure~\ref{fig:TheoCMDs} it can be seen that, for a 25\,Myr old cluster (log($t$) $ = 7.4$) with 1,000\,$M_\odot$, missing only a couple of bright members near the MSTO of the cluster might easily result in an age determination offset of 75\,Myr (0.6 in logarithmic scale). Therefore, for the younger clusters in our sample, large age uncertainties are expected to be found, making them inadequate as observable checks of the PA relation of Cepheids. 

\par In this regard, since the initial mass functions are sampled stochastically, fewer stars are formed overall in less massive clusters, and as a consequence, the probability of forming intermediate and high-mass stars decreases, especially when the star formation rate of a cluster is rather low \citep[see, e.g.,][]{Weidner06,Eldridge12}.
In fact, the stochastic sampling that affects the observed mock cluster colour-magnitude diagrams coupled with the relatively young age of the clusters, in addition to the natural limitations of the Gaia photometry \citep[saturation, for instance, and crowding for distant clusters;][]{Boubert20}, translates into not well defined stellar main-sequence turn-offs and an evident dearth of cluster members at the turn-offs and post-main-sequence evolutionary stages. For more details regarding the effects of stochasticity in cluster populations we refer the reader to more dedicated works, such as those carried out by \citet{Fouesneau10} and \citet{Popescu10}. 

\par As a final remark, we would like to highlight that in order to overcome the limitation described above, extremely high levels of both purity and completeness in the open cluster member catalogues are mandatory for an age determination at the accuracy required (a few tens of Myr) to constrain Cepheid PA relations. In particular, no turn-off or further evolved members should be missing or falsely included. The use of reliable spectral types, in addition to high-precision reddening estimates should become beneficial to better constrain the cluster ages. If these requirements are not met, only such clusters with well-defined MSTO can in practice be used with confidence, which biases the cluster Cepheid sample selection towards older and more massive hosts. Similar conclusions were reached by \citet{Senchyna15} in their study of the PA relation of Cepheids in M31, in which they attribute the broad constraints in their PA fits to the difficulty of assigning an age to low-mass clusters at large distances.

\par This is a reason why, from all the high probability combos obtained in Section~\ref{sec:outcome}, we retain only 11 cluster-Cepheid pairs in an attempt to characterize Cepheids' PA relations in the next section.
\par 

\begin{figure*}
\begin{center}
\includegraphics[angle=0,scale=.5]{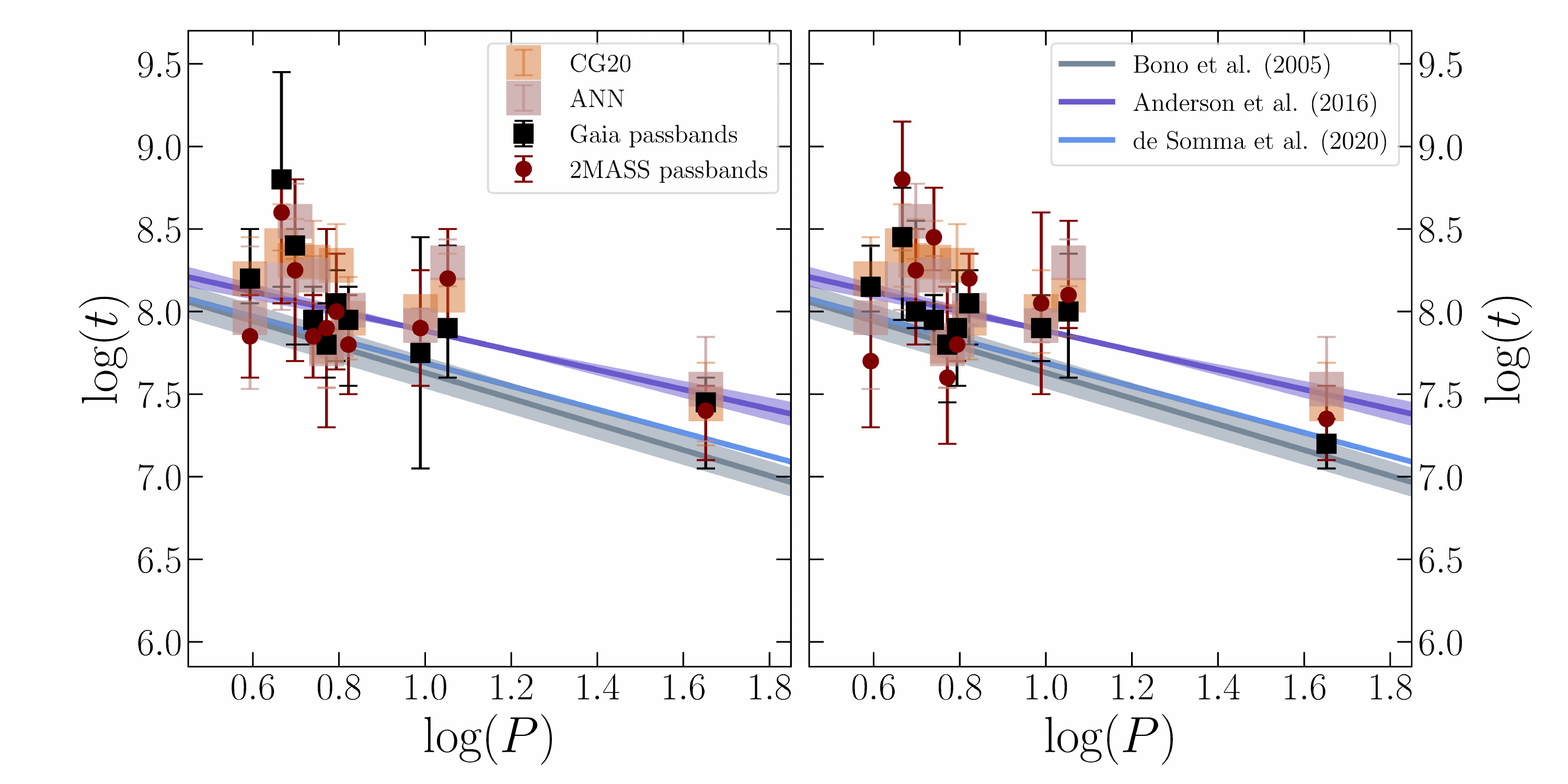}
\caption{Comparison between the logarithm of the Cepheid periods in days \citep{Udalski18,Ripepi19} and the cluster logarithmic ages (in years), obtained with different isochrone sets: PARSEC (left panel) and MIST (right panel) for the combos shown in Table~\ref{tab:ages}. We do not include in the plot the unique first overtone Cepheid recovered as high probability cluster member. Black symbols represent the cluster ages obtained using Gaia DR2 photometry and red stars those obtained using the 2MASS passbands. For comparison, we overplot these values to the theoretical PA relation for fundamental-mode Cepheids derived by \citet{Bono05}, in gray, \citet{Anderson16}, in blue, and \citet{deSomma20b} in light blue, for $Z = 0.010$, $Z=0.014$, and $Z = 0.020$, respectively.
The cluster ages obtained by CG20 and by our ANN analysis are also shown as a reference, setting a constant uncertainty budget at 0.25\,dex for the former.
The errors on the period are negligible and therefore not included in the plots.}
\label{fig:PAs}
\end{center}
\end{figure*}

\subsection{The Cepheid Period-Age Relation}
\label{sec:PAs}

\par
\citet{Bono05} provided the first theoretical PA relations for Cepheids with Magellanic or solar-like chemical compositions. More recently, models including rotation have been developed \citep{Anderson14,Anderson16} predicting that, as rotation increases the main-sequence lifetime of the stars, higher Cepheid ages are expected in comparison with ages determined using non-rotating models (by $\Delta$ log($t$) $\sim  0.2$ to $0.3$). Taking advantage of updated evolutionary \citep{Hidalgo18} and pulsation \citep{deSomma20a} models, \citet{deSomma20b} derived new period-age relations and period-age-colour relations in the Gaia passbands.

\par It is natural to overlay the age of a cluster Cepheid as provided by the determination of the age of the hosting cluster (following the traditional assumption that they are coeval) on the prediction of the Cepheids' age as given by a theoretical PA relation. As described above, we take into account only those (11) clusters for which we consider the ages relatively well constrained. Considering this small number together with their bias towards larger ages, we do not consider fitting an empirical PA relation. For the same reason, we only consider fundamental mode Cepheids, since only a single first overtone Cepheid falls in this restricted sample.

\par The comparison with the theoretical PA relations of \citet{Bono05} (no stellar rotation), \citet{Anderson16} ($\Omega_{\rm ZAMS}/\Omega_{\rm crit} = 0.5$), and \citet{deSomma20b} (no rotation) is displayed in Figure~\ref{fig:PAs}. Assuming a PA relation of the shape log($t$) = $\alpha + \beta \cdot $log($P$), where the value of the age $t$ is represented in years, the former study reports $\alpha = 8.41 \pm 0.10$ and $\beta = -0.78 \pm 0.01$ for fundamental-mode Cepheids and for $Z=0.01$ \citep[Table 4 in][]{Bono05}. In the case of \citet{Anderson16} we used the average PA relation slope and intercept for a Cepheid with $Z = 0.014$ in their second and third crossing of the instability strip \citep[Table 4 in][]{Anderson16}, where $\alpha = 8.48 \pm 0.09$ and $\beta = -0.59 \pm 0.09$ for fundamental mode Cepheids. The uncertainty assumed represents the standard deviation of these values with respect to the averages. Finally, for the models of \citet{deSomma20b} we adopt the results obtained in the case of a canonical mass-luminosity relation, when neglecting rotation, mass-loss, and overshooting, with a mixing length parameter equal to 1.5, and $Z = 0.020$ \citep[Table 2 in][]{deSomma20b}. In this case $\alpha = 8.39 \pm 0.01$ and $\beta = -0.70 \pm 0.01$ for fundamental-mode Cepheids. 

\par From the plots shown in Figure~\ref{fig:PAs}, we do not observe well-defined relations between the Cepheid periods and the cluster ages, regardless of the choice of photometric system, isochrone models (with and without stellar rotation), or source of the cluster ages (from the use of the ANN, or from CG20). The large scatter of the cluster ages as compared to the theoretical predictions for the Cepheids' ages reveals the lack of accuracy and precision of our age determinations for this purpose, and makes it impossible to discriminate between the various theoretical Cepheid PA relations. Although not a factor here, we stress that a proper comparison between theoretical and empirical ages should be carried out with models making the same assumptions on overshooting, rotation, etc.

\section{Discussion and concluding remarks}
\label{sec:conclusions}

\par Considering that the number of known open clusters and classical Cepheids has increased considerably in recent years, and taking advantage of the unprecedented quality of the data provided by the Gaia mission (DR2 and eDR3), we revisited the membership of classical Cepheids in Galactic open clusters. We follow the Bayesian approach proposed by \citet{Anderson13}, focusing only on the relative position and kinematics of the Cepheid and its potential host.  

\par After investigating more than $40,000$ possible combinations (combos) selected by their on-sky projected distance, we found 67 with a probability of association larger than 10\,per cent, including 44 with a posterior probability larger than 25\,per cent.
Additionally, we found 96 possible associations with probabilities between 1 and 10\,per cent, mostly in newly discovered open clusters. 
Within the list of combos with probabilities $>$ 0.01 (163 in total), we report 19 that are consistent with previously known cluster Cepheids.
Six literature combos are unlikely associations given their extremely small membership probabilities. 

\par We advocate for dedicated follow-up studies including a detailed chemical investigation \citep["chemical tagging";][]{Freeman02} and an accurate age determination. For combos with a membership probability higher than 10\,per cent, logarithmic ages range from 6.42 (2.6\,Myr) and 8.72 (525\,Myr), with a median age of 7.80 (63\,Myr), according to the age determinations of \citet{CG20b}.

\par In an attempt to compare the age of Cepheids as given through isochrone fitting of the cluster population with the age given by theoretical  period-age relations, we derived cluster ages using our own method and a publicly available code based on artificial neural networks. Despite an overall good agreement with literature values, we conclude that current age determinations for young open clusters do not reach the required accuracy (log($t$)~$<$ 0.2) for the proposed goal. We argue that the reason is intrinsic to young open clusters, especially the less massive ones, due to the lack of MSTO stars. Such conclusions have already been reported by, e.g., \citet{Senchyna15}. We believe that upcoming Gaia data releases will allow us to overcome this difficulty by providing colour-magnitude diagrams with extremely high levels of completeness and purity. Other approaches to possibly avoid some of these difficulties could come from the comparison of their observed and theoretical luminosity functions \citep[see, e.g.,][]{Piskunov04}, or by complementing Gaia data by reddening-free indices and spectral types for upper main-sequence stars. 
We note in passing that \citet{PenaRamirez21} pre-selected potential cluster members via their Gaia DR2 proper motions using Gaussian mixture models, and assigned membership probabilities using the same unsupervised machine learning method (UPMASK) as in \citet{CG18b} and \citet{CG20a}, but on near-infrared data instead of Gaia DR2 photometry. For the six clusters in their study, they report on average 45\,per cent more cluster members than \citet{CG18b} and \citet{CG20a}. 

\par Despite a much larger number of clusters and Cepheids in the input catalogues, the number of high probability combos did not increase much. Even assuming that all cluster Cepheids pairs with $P(A|B) > 0.01$ represent a true association, which is far from being realistic, we infer 4.1\,per cent (121/2921)\footnote{For this, we also impose that a given Cepheid is associated with a unique cluster, therefore we consider 121 combos instead of 163.} as an upper limit to the fraction of classical Cepheids in open clusters. \citet{Anderson18}, using the bona-fide cluster Cepheids described in A13, report an  upper limit of 8.5\,per cent for the clustered fraction of fundamental-mode Cepheids within 2\,kpc from the Sun. \citet{Anderson18} also estimated the clustered fraction of fundamental-mode Cepheids in the Small and Large Magellanic Clouds to 6 and 11\,per cent, respectively, and to 2.5\,per cent in M31. Although the fraction presumably varies from galaxy to galaxy, even within a given galaxy, and also with time, these numbers all suggest a low fraction of Cepheids in clusters. 

\par Since very young ($<$ 20\,Myr) clusters are overabundant, it is known that young clusters dissolve quickly, whether it is a consequence of gas expulsion or of their stellar dynamic and stellar evolution-driven expansion \citep[e.g.,][]{Lada03,Goodwin06,Moeckel12}, and these dissolutions can occur as fast as 100\,Myr even for relatively massive clusters \citep[of $\sim$1,000\,$M_\odot$][]{Shukirgaliyev18}. In contrast, Cepheids have ages ranging from a few tens to a few hundred Myr \citep[e.g.,][]{Anderson16}. The low fraction of Cepheids in clusters could then be related to the rapid dissolution of young clusters, or alternatively indicate that they are born elsewhere.

\par After estimating the age distribution of Galactic open clusters within a cylinder of 2~kpc radius from the Sun (taking into account age-dependent completeness limits), \citet{Anders20} derived the star formation rate in clusters in the Solar vicinity, compared it to the total star formation rate in the solar vicinity \citep{Mor19}, and concluded that only $\sim$16\,per cent of stars formed in bound clusters. This result is in line with recent findings where star formation takes place at all scales and in unbound structures rather than in clusters (see e.g., \citealt{Ward20}, or the review by \citealt{Adamo20}). \\

\par To better investigate the birthplace of Cepheids, one could envision adapting the prior (which currently only relies on the on-sky separation between clusters and Cepheids) to take into account the dynamical state of the cluster (its possible ongoing dissolution), in line with the findings of, e.g., \citet{Meingast21}. This could already provide higher membership probabilities for Cepheids about to leave a cluster, as it seems to be the case for QZ~Nor as reported by \cite{Breuval20}, or those who just left it. An even more promising alternative would be to exploit conservative integrals of motion, as is done in the halo to identify globular cluster escapees and link them to their original cluster \citep[see e.g.,][]{Hanke20}. Together with the aforementioned chemical tagging, this might render it possible to associate a given Cepheid and other nearby young stars with a unique birthplace, of any nature whatsoever.

\section*{Acknowledgements}
We thank the anonymous referee for her/his thorough feedback, which helped improve the quality of this manuscript. 
The results of this work were obtained using data from the European Space Agency (ESA) mission Gaia (https://www.cosmos.esa.int/gaia), processed by  the Gaia Data Processing and Analysis Consortium  (DPAC). 
Funding for the DPAC has been provided by  national institutions, in particular the institutions participating in the Gaia Multilateral Agreement (MLA).
G.M. acknowledges S. Yen and M. Hanke for fruitful discussions and insights regarding the data analysis used for this study.
G.M. gratefully acknowledges the support of the Hector Fellow Academy.
B.L. and E.K.G. acknowledge the Deutsche Forschungsgemeinschaft (DFG, German Research Foundation) -- Project-ID 138713538 -- SFB 881 (“The Milky Way System”, subprojects A05, B05).
This research has made use of pandas \citep{McKinney10}, numpy \citep{vanderWalt11}, the Astropy library \citep{Astropy13,Astropy18}, and the software TOPCAT \citep{Taylor05}. 
This research has made use of the VizieR catalogue access tool, CDS, Strasbourg, France. 
The original description of the VizieR service was published in A\&AS 143, 23.
The figures in this paper were produced with Matplotlib \citep{Hunter07}.

\section*{Data Availability}

The data underlying this article are publicly available and their sources are described in detail in this document. Tables containing the results of this study are provided as online supplementary material.




\appendix

\renewcommand{\thefigure}{A\arabic{figure}}
\setcounter{figure}{0}

\renewcommand{\thetable}{A\arabic{table}}
\setcounter{table}{0}

\onecolumn
\LTcapwidth=\textwidth
\begin{longtable}[c]{cccHcccccH}
 \caption{Cluster - Cepheid pairs with membership probabilities $P(A|B)$ between 0.01 and 0.10. The table lists the cluster names as well as their MWSC identification in the K13 catalogue, the Cepheid names, the angular separation of the pair over the cluster's $r_1$ (Sep/$r_1$), the list of constraints used to derive the membership probability, the prior $P(A)$, the likelihood $P(A|B)$ and the membership probability $P(B|A)$. The full table is provided as supplementary material.  \label{tab:combosA2}}\\
 \hline
 Open cluster & MWSC ID & Cepheid & RUWE & $\rm{Sep}/r_1$ & Constraints & $P(A)$ & $P(B|A)$ & $P(A|B)$ & Ref.\\
 \hline
 \endfirsthead

 Continuation of Table \ref{tab:combosA2}\\
 \hline
 Open cluster & MWSC ID & Cepheid & RUWE & $\rm{Sep}/r_1$ & Constraints & $P(A)$ & $P(B|A)$ & $P(A|B)$ & Ref.\\
 \hline
 \endhead

 \hline
 \endfoot

 \hline
 \endlastfoot
       UFMG~54 &    -- &              OGLE-GD-CEP-0964 &   1.13 &  11.56 &         $\varpi$, $\mu_{\alpha}^{*}$, $\mu_{\delta}$ &  0.10 &  1.00 &  0.10 &  -- \\
      NGC~6664 &  2962 &                        EV~Sct &   1.01 &   3.64 &  $\varpi$, $V_r$, $\mu_{\alpha}^{*}$, $\mu_{\delta}$ &  0.12 &  0.80 &  0.10 &  -- \\
        LP~888 &    -- &                        CG~Cas &   1.03 &   2.69 &         $\varpi$, $\mu_{\alpha}^{*}$, $\mu_{\delta}$ &  0.27 &  0.35 &  0.09 &  -- \\
         LP~58 &    -- &                        AM~Vel &  41.08 &   4.09 &         $\varpi$, $\mu_{\alpha}^{*}$, $\mu_{\delta}$ &  0.09 &  1.00 &  0.09 &  -- \\
         BH~99 &  1831 &              OGLE-GD-CEP-1669 &  12.61 &   1.68 &         $\varpi$, $\mu_{\alpha}^{*}$, $\mu_{\delta}$ &  0.58 &  0.16 &  0.09 &  -- \\
     Platais~8 &  1629 &              OGLE-GD-CEP-0507 &  26.32 &   3.59 &         $\varpi$, $\mu_{\alpha}^{*}$, $\mu_{\delta}$ &  0.13 &  0.66 &  0.09 &  -- \\
       UBC~135 &    -- &                        GI~Cyg &   1.01 &   0.98 &         $\varpi$, $\mu_{\alpha}^{*}$, $\mu_{\delta}$ &  1.00 &  0.08 &  0.08 &  -- \\
 Collinder~228 &  1845 &           GDS~J1046447-601605 &   0.97 &   2.76 &         $\varpi$, $\mu_{\alpha}^{*}$, $\mu_{\delta}$ &  0.26 &  0.32 &  0.08 &  -- \\
    Loden~1409 &  2249 &              OGLE-GD-CEP-0998 &   6.82 &   3.46 &         $\varpi$, $\mu_{\alpha}^{*}$, $\mu_{\delta}$ &  0.10 &  0.81 &  0.08 &  -- \\
      NGC~5662 &  2234 &                         V~Cen &   1.06 &   1.51 &  $\varpi$, $V_r$, $\mu_{\alpha}^{*}$, $\mu_{\delta}$ &  0.67 &  0.12 &  0.08 &  -- \\

\end{longtable}
{\raggedright 
    
     \vspace{0.001cm}
     {\raggedright 
     {\scriptsize $^*$ Uncertain Cepheid classification, as noted by the OGLE team. }\\
     {\scriptsize $^+$ Coincidence with a cluster in the catalogues of \citet{Liu19} or \citet{Sim19}, according to \citet{Castro-Ginard20}. }\\ 
      \par}
      \par}
\twocolumn
\vspace{0.001cm}

\begin{table*}

\caption{Prior, likelihood and membership probability for literature combos. We list the cluster and Cepheid names, their angular separation as a function of $r_1$, and the constraints used in the analysis. 
For the combos in the top list, which appear at least once in 
Turner's database (only potential combos with open clusters), A13 (with $P(A|B)$~$>$ 0.10, from that work), or \citet{Chen15}, we obtain membership probabilities $>$1\,per cent. We included in this list the cluster Cepheids recently confirmed by \citet{Clark15}, \citet{Lohr18}, and \citet{Negueruela20}, as we considered them in the combo descriptions presented in Section~\ref{sec:outcome}.  
Combos in the bottom part of this table are those considered missed, and are listed as true associations in at least three of the aforementioned catalogues plus \citet{Roeck12}.
In the last column we list references where the Cepheid membership to the cluster is discussed: 
a :  \citet{Lohr18}, 
b :  \citet{Negueruela18}, 
c :  \citet{Anderson13}, 
d :  \citet{Walker85a}, 
e :  \citet{An07}, 
f :  \citet{Turner10}, 
g :  \citet{Majaess11}, 
h :  \citet{Chen15}, 
i :  \citet{Sandage58}, 
j :  \citet{Mateo88}, 
k :  \citet{Matthews95}, 
l :  \citet{Irwin1955}, 
m :  \citet{Kholopov56}, 
n :  \citet{Feast57}, 
o :  \citet{Pel85}, 
p :  \citet{Turner92}, 
q :  \citet{Turner98c}, 
r :  \citet{Coulson85}, 
s :  \citet{Walker85b}, 
t :  \citet{Hoyle03}, 
u :  \citet{Clark15}, 
v :  \citet{Flower78}, 
w :  \citet{Turner81}, 
x :  \citet{Negueruela20}, 
y :  \citet{Turner86}, 
z :  \citet{Walker87}, 
$\rm{\alpha}$ :  \citet{Turner76}, 
$\rm{\beta}$ :  \citet{Schmidt82}, 
$\rm{\gamma}$ :  \citet{Turner02}, 
$\rm{\delta}$ :  \citet{Turner82}, 
$\rm{\epsilon}$ :  \citet{Claria91}, 
$\rm{\zeta}$ :  \citet{Turner08}, 
$\rm{\eta}$ :  \citet{Turner85}, 
$\rm{\theta}$ :  \citet{Turner98a}, 
$\rm{\iota}$ :  \citet{Yilmaz66}, 
$\rm{\kappa}$ :  \citet{Turner80}, 
$\rm{\iota}$ :  \citet{Turner93}, 
$\rm{\lambda}$ :  \citet{Turner98b}. 
}

\begin{center}
\label{tab:combosA1}
{
\scriptsize

\begin{supertabular}{cHcHcccccc}  
\multicolumn{10}{c}{\textbf{Bona fide combos recovered}}\\
\noalign{\smallskip}
\hline
Open cluster & MWSC & Cepheid & RUWE & $\rm{Sep}/r_1$ & Constraints & $P(A)$ & $P(B|A)$ & $P(A|B)$ & References  \\
\noalign{\smallskip}
\hline

 Berkeley~55 &  3490 &  ASASSN-V~J211659.90+514558.7 &  1.09 &   0.29 &         $\varpi$, $\mu_{\alpha}^{*}$, $\mu_{\delta}$ &  1.00 &  0.94 &  0.94 &                                           a \\
 Berkeley~51 &  3280 &  ASASSN-V~J201151.18+342447.2 &  1.07 &   0.75 &         $\varpi$, $\mu_{\alpha}^{*}$, $\mu_{\delta}$ &  1.00 &  0.85 &  0.85 &                                        a, b \\
     Lynga~6 &  2348 &                        TW~Nor &  0.89 &   0.39 &  $\varpi$, $V_r$, $\mu_{\alpha}^{*}$, $\mu_{\delta}$ &  1.00 &  0.82 &  0.82 &                               c, d, e, f, g \\
    NGC~7790 &  3781 &                        CF~Cas &  1.04 &   0.30 &  $\varpi$, $V_r$, $\mu_{\alpha}^{*}$, $\mu_{\delta}$ &  1.00 &  0.80 &  0.80 &                               c, h, i, j, k \\
     IC~4725 &  2940 &                         U~Sgr &  0.85 &   0.10 &  $\varpi$, $V_r$, $\mu_{\alpha}^{*}$, $\mu_{\delta}$ &  1.00 &  0.75 &  0.75 &                         c, e, h, l, m, n, o \\
     NGC~129 &    53 &                        DL~Cas &  1.88 &   0.04 &  $\varpi$, $V_r$, $\mu_{\alpha}^{*}$, $\mu_{\delta}$ &  1.00 &  0.75 &  0.75 &                                  c, h, m, p \\
   vdBergh~1 &   934 &                        CV~Mon &  1.10 &   0.68 &         $\varpi$, $\mu_{\alpha}^{*}$, $\mu_{\delta}$ &  1.00 &  0.67 &  0.67 &                                     c, h, q \\
    NGC~6067 &  2370 &                     V0340~Nor &  0.92 &   0.20 &         $\varpi$, $\mu_{\alpha}^{*}$, $\mu_{\delta}$ &  1.00 &  0.66 &  0.66 &                            c, e, f, r, s, t \\
      BH~222 &  2564 &              OGLE-BLG-CEP-110 &  0.98 &   0.41 &         $\varpi$, $\mu_{\alpha}^{*}$, $\mu_{\delta}$ &  1.00 &  0.65 &  0.65 &                                        a, u \\
    NGC~6649 &  2949 &                     V0367~Sct &  1.02 &   0.73 &         $\varpi$, $\mu_{\alpha}^{*}$, $\mu_{\delta}$ &  1.00 &  0.63 &  0.63 &                                     c, v, w \\
     UBC~130 &    -- &                        SV~Vul &  1.20 &   1.41 &         $\varpi$, $\mu_{\alpha}^{*}$, $\mu_{\delta}$ &  0.73 &  0.71 &  0.52 &                                           x \\
    NGC~7790 &  3781 &                      CE~Cas~B &  1.20 &   0.51 &         $\varpi$, $\mu_{\alpha}^{*}$, $\mu_{\delta}$ &  1.00 &  0.50 &  0.50 &                               c, h, i, j, k \\
    FSR~0158 &  3182 &                        GQ~Vul &  0.91 &   1.80 &         $\varpi$, $\mu_{\alpha}^{*}$, $\mu_{\delta}$ &  0.53 &  0.89 &  0.47 &                                           c \\
    NGC~6087 &  2382 &                         S~Nor &  0.88 &   0.14 &  $\varpi$, $V_r$, $\mu_{\alpha}^{*}$, $\mu_{\delta}$ &  1.00 &  0.38 &  0.38 &                            c, h, l, m, n, y \\
    NGC~7790 &  3781 &                      CE~Cas~A &  1.09 &   0.50 &         $\varpi$, $\mu_{\alpha}^{*}$, $\mu_{\delta}$ &  1.00 &  0.28 &  0.28 &                               c, h, i, j, k \\
 Ruprecht~79 &  1701 &                        CS~Vel &  0.91 &   0.84 &  $\varpi$, $V_r$, $\mu_{\alpha}^{*}$, $\mu_{\delta}$ &  1.00 &  0.23 &  0.23 &                                        f, z \\
    NGC~6664 &  2962 &                        EV~Sct &  1.01 &   3.64 &  $\varpi$, $V_r$, $\mu_{\alpha}^{*}$, $\mu_{\delta}$ &  0.12 &  0.80 &  0.10 &  $\rm{\alpha}$, $\rm{\beta}$, $\rm{\gamma}$ \\
    NGC~5662 &  2234 &                         V~Cen &  1.06 &   1.51 &  $\varpi$, $V_r$, $\mu_{\alpha}^{*}$, $\mu_{\delta}$ &  0.67 &  0.12 &  0.08 &     e, f, h, $\rm{\delta}$, $\rm{\epsilon}$ \\
     ASCC~69 &  1996 &                         S~Mus &  4.49 &   4.66 &  $\varpi$, $V_r$, $\mu_{\alpha}^{*}$, $\mu_{\delta}$ &  0.06 &  0.58 &  0.03 &                                        c, h \\

\noalign{\smallskip}
\hline

\noalign{\bigskip}
\multicolumn{10}{c}{\textbf{Missed}}\\
\noalign{\smallskip}
\hline
Open cluster & MWSC & Cepheid & RUWE & $\rm{Sep}/r_1$ & Constraints & $P(A)$ & $P(B|A)$ & $P(A|B)$ & References  \\
\noalign{\smallskip}
\hline

   Berkeley~58 &     1.0 &  CG~Cas &  1.03 &   1.60 &  $\varpi$, $V_r$, $\mu_{\alpha}^{*}$, $\mu_{\delta}$ &  0.62 &  0.00 &  0.00 &                       h, $\rm{\zeta}$ \\
 Collinder~394 &  3011.0 &  BB~Sgr &  0.82 &   1.59 &  $\varpi$, $V_r$, $\mu_{\alpha}^{*}$, $\mu_{\delta}$ &  0.63 &  0.00 &  0.00 &                  c, f, h, $\rm{\eta}$ \\
  Ruprecht~175 &  3409.0 &   X~Cyg &  1.28 &   7.52 &         $\varpi$, $\mu_{\alpha}^{*}$, $\mu_{\delta}$ &  0.00 &  0.00 &  0.00 &                      h, $\rm{\theta}$ \\
   Trumpler~35 &  2980.0 &  RU~Sct &  0.87 &   5.12 &  $\varpi$, $V_r$, $\mu_{\alpha}^{*}$, $\mu_{\delta}$ &  0.04 &  0.00 &  0.00 &  c, f, s, $\rm{\iota}$, $\rm{\kappa}$ \\
      Turner~2 &  2872.0 &  WZ~Sgr &  0.94 &   2.56 &         $\varpi$, $\mu_{\alpha}^{*}$, $\mu_{\delta}$ &  0.21 &  0.00 &  0.00 &                    c, h, $\rm{\iota}$ \\
      Turner~9 &  3168.0 &  SU~Cyg &  3.44 &   0.01 &  $\varpi$, $V_r$, $\mu_{\alpha}^{*}$, $\mu_{\delta}$ &  1.00 &  0.00 &  0.00 &                  c, f, $\rm{\lambda}$ \\

   Berkeley~58 &     1.0 &  CG~Cas &  1.03 &   1.60 &  $\varpi$, $V_r$, $\mu_{\alpha}^{*}$, $\mu_{\delta}$ &  0.62 &  0.00 &  0.00 &                       h, $\rm{\zeta}$ \\
 Collinder~394 &  3011.0 &  BB~Sgr &  0.82 &   1.59 &  $\varpi$, $V_r$, $\mu_{\alpha}^{*}$, $\mu_{\delta}$ &  0.63 &  0.00 &  0.00 &                  c, f, h, $\rm{\eta}$ \\
  Ruprecht~175 &  3409.0 &   X~Cyg &  1.28 &   7.52 &         $\varpi$, $\mu_{\alpha}^{*}$, $\mu_{\delta}$ &  0.00 &  0.00 &  0.00 &                      h, $\rm{\theta}$ \\
   Trumpler~35 &  2980.0 &  RU~Sct &  0.87 &   5.12 &  $\varpi$, $V_r$, $\mu_{\alpha}^{*}$, $\mu_{\delta}$ &  0.04 &  0.00 &  0.00 &  c, f, t, $\rm{\iota}$, $\rm{\kappa}$ \\
      Turner~2 &  2872.0 &  WZ~Sgr &  0.94 &   2.56 &         $\varpi$, $\mu_{\alpha}^{*}$, $\mu_{\delta}$ &  0.21 &  0.00 &  0.00 &                    c, h, $\rm{\iota}$\\
      Turner~9 &  3168.0 &  SU~Cyg &  3.44 &   0.01 &  $\varpi$, $V_r$, $\mu_{\alpha}^{*}$, $\mu_{\delta}$ &  1.00 &  0.00 &  0.00 &                  c, f, $\rm{\lambda}$ \\

\noalign{\smallskip}
\hline

\end{supertabular}
}
\end{center}
\end{table*}

\onecolumn
\LTcapwidth=\textwidth
\begin{longtable}[c]{cccHccHcHH}
 \caption{Same as Table~\ref{tab:combosA2} for combos  
 Cluster - Cepheid pairs with low priors $P(A)$, but high likelihood ($P(B|A)>0.85$). The full table is provided as supplementary material.  \label{tab:combosA3}}\\
 \hline
 Open cluster & MWSC ID & Cepheid & RUWE & $\rm{Sep}/r_1$ & Constraints & $P(A)$ & $P(B|A)$ & $P(A|B)$ & Ref.\\
 \hline
 \endfirsthead

 Continuation of Table \ref{tab:combosA3}\\
 \hline
 Open cluster & MWSC ID & Cepheid & RUWE & $\rm{Sep}/r_1$ & Constraints & $P(A)$ & $P(B|A)$ & $P(A|B)$ & Ref.\\
 \hline
 \endhead

 \hline
 \endfoot

 \hline
 \endlastfoot
           ASCC~59 &  1793 &              OGLE-GD-CEP-0507 &  26.32 &    7.38 &         $\varpi$, $\mu_{\alpha}^{*}$, $\mu_{\delta}$ &  0.00 &  0.99 &  0.00 &  -- \\
          ASCC~60 &  1823 &              OGLE-GD-CEP-0507 &  26.32 &   12.12 &         $\varpi$, $\mu_{\alpha}^{*}$, $\mu_{\delta}$ &  0.00 &  0.86 &  0.00 &  -- \\
        Alessi~18 &  1284 &                        BE~Pup &   1.45 &   51.21 &         $\varpi$, $\mu_{\alpha}^{*}$, $\mu_{\delta}$ &  0.00 &  0.90 &  0.00 &  -- \\
           BH~118 &  1939 &              OGLE-GD-CEP-0633 &   1.01 &   76.23 &         $\varpi$, $\mu_{\alpha}^{*}$, $\mu_{\delta}$ &  0.00 &  0.87 &  0.00 &  -- \\
           BH~144 &  2098 &              OGLE-GD-CEP-0869 &   1.04 &   49.27 &         $\varpi$, $\mu_{\alpha}^{*}$, $\mu_{\delta}$ &  0.00 &  1.00 &  0.00 &  -- \\
           BH~144 &  2098 &              OGLE-GD-CEP-0873 &   0.94 &   50.10 &         $\varpi$, $\mu_{\alpha}^{*}$, $\mu_{\delta}$ &  0.00 &  0.98 &  0.00 &  -- \\
           BH~144 &  2098 &           GDS~J1313507-642626 &   1.31 &   61.72 &         $\varpi$, $\mu_{\alpha}^{*}$, $\mu_{\delta}$ &  0.00 &  0.94 &  0.00 &  -- \\
           BH~151 &  2147 &         WISE~J132924.6-625511 &   1.16 &  215.66 &         $\varpi$, $\mu_{\alpha}^{*}$, $\mu_{\delta}$ &  0.00 &  0.85 &  0.00 &  -- \\
           BH~222 &  2564 &              OGLE-BLG-CEP-173 &   0.98 &   85.59 &         $\varpi$, $\mu_{\alpha}^{*}$, $\mu_{\delta}$ &  0.00 &  0.93 &  0.00 &  -- \\
            BH~63 &  1653 &              OGLE-GD-CEP-0328 &   0.98 &   23.97 &         $\varpi$, $\mu_{\alpha}^{*}$, $\mu_{\delta}$ &  0.00 &  0.99 &  0.00 &  -- \\

\end{longtable}

{\raggedright 
    
     \vspace{0.001cm}
     {\raggedright 
     {\scriptsize $^k$ Cluster with position and proper motion compatible with a cluster in K13, based on \citet{Castro-Ginard20}. }\\
      \par}
      \par}

\twocolumn

\begin{figure*}

\includegraphics[angle=0,scale=.29]{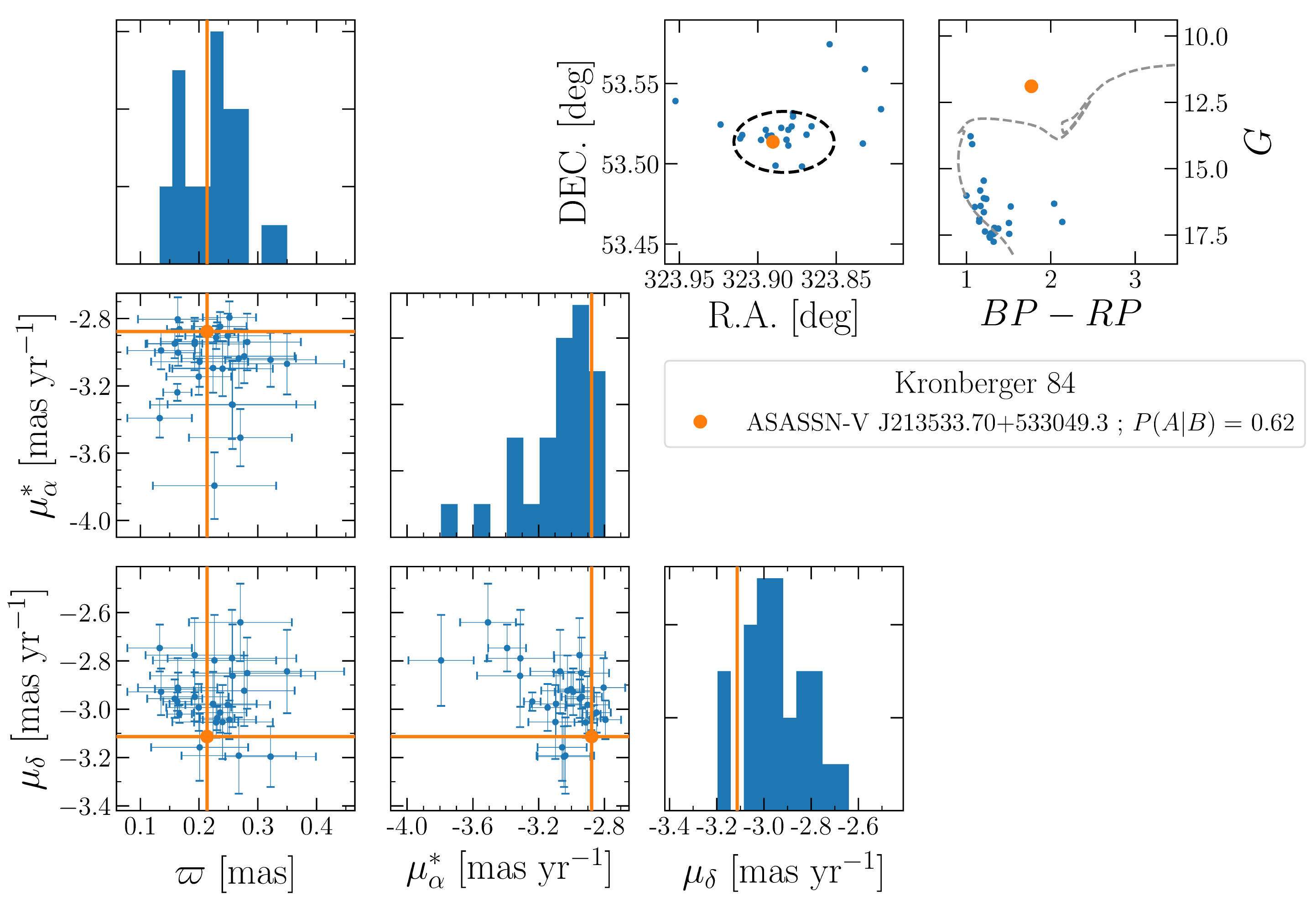}
\includegraphics[angle=0,scale=.29]{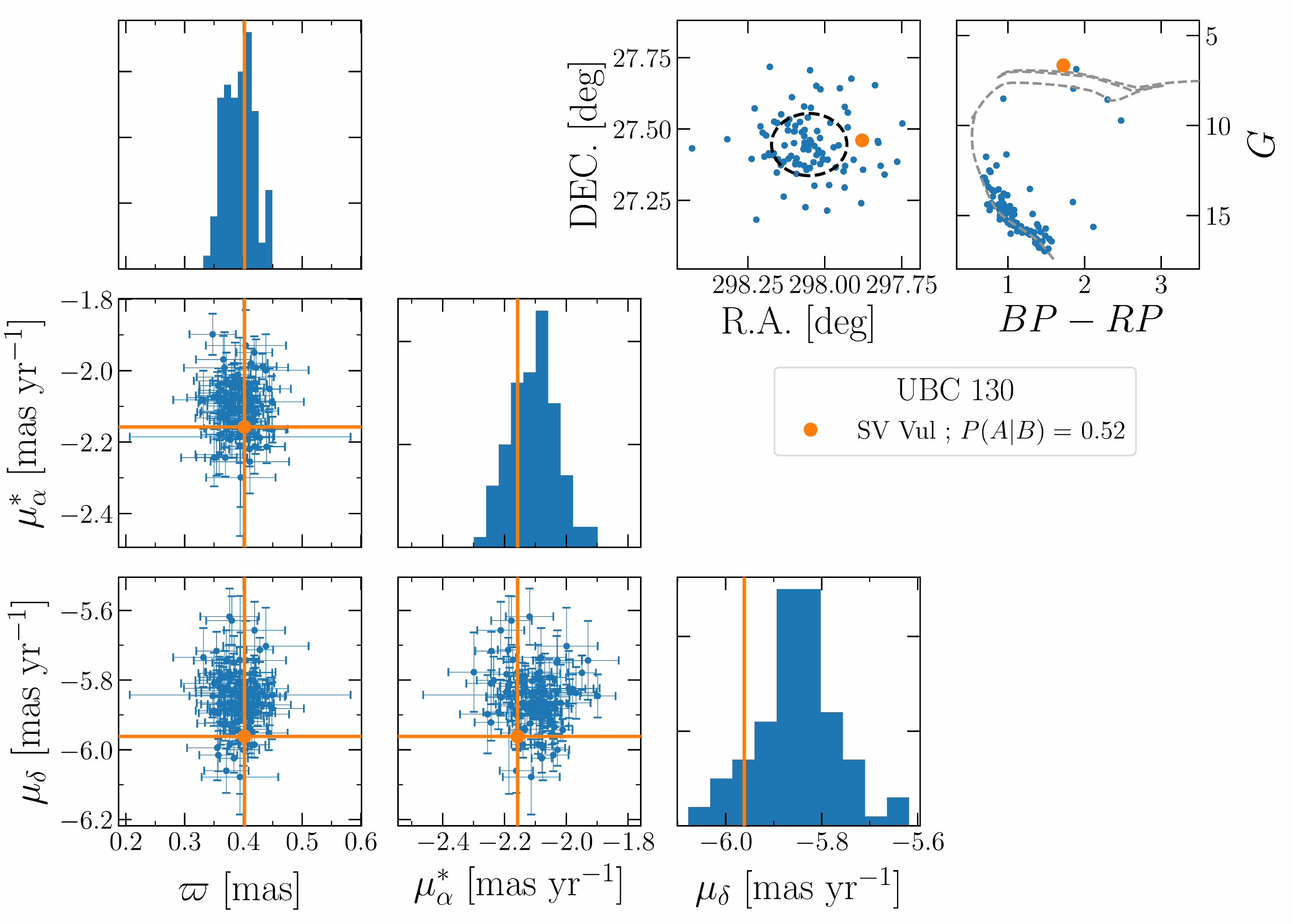}
\includegraphics[angle=0,scale=.29]{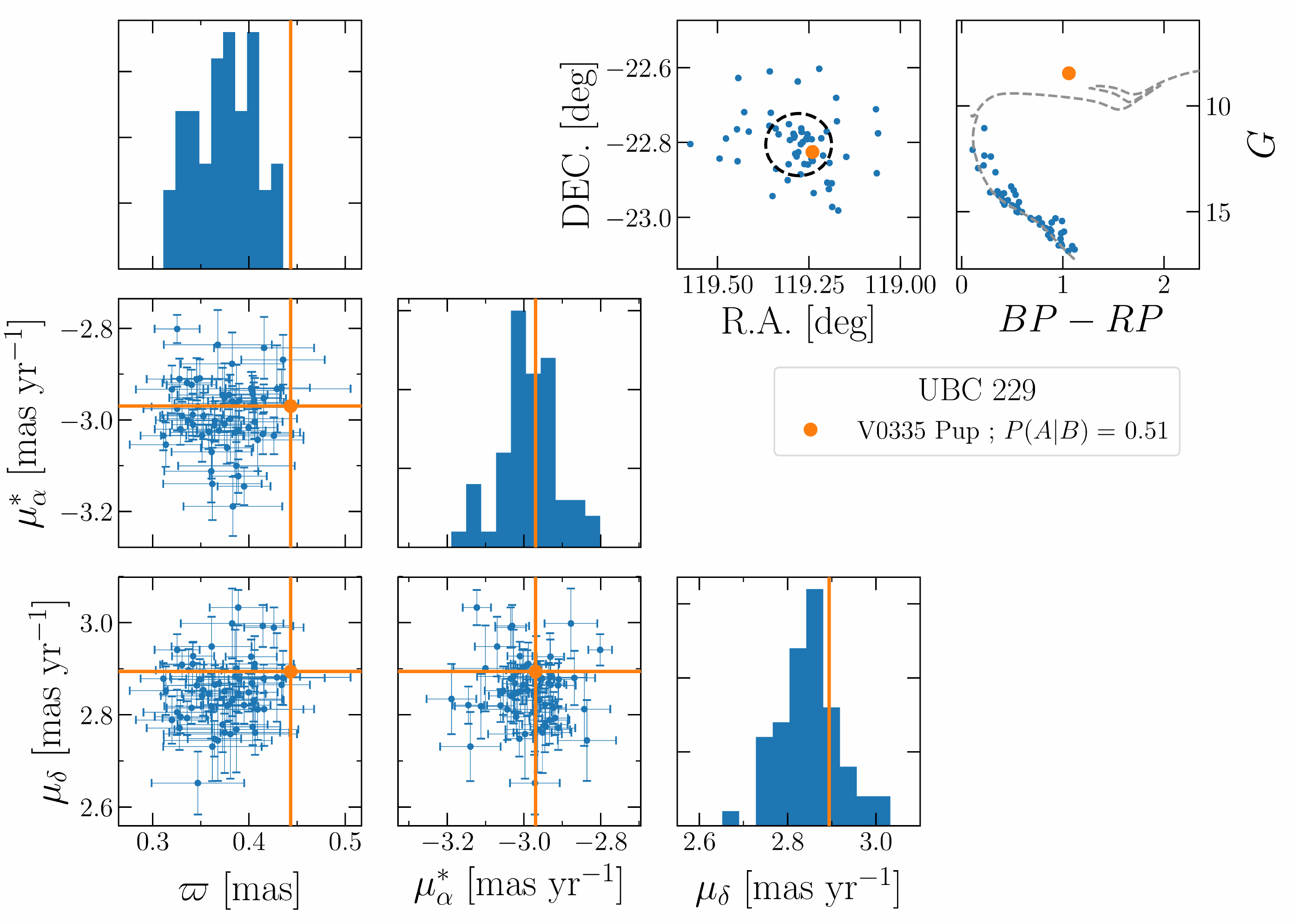}
\includegraphics[angle=0,scale=.29]{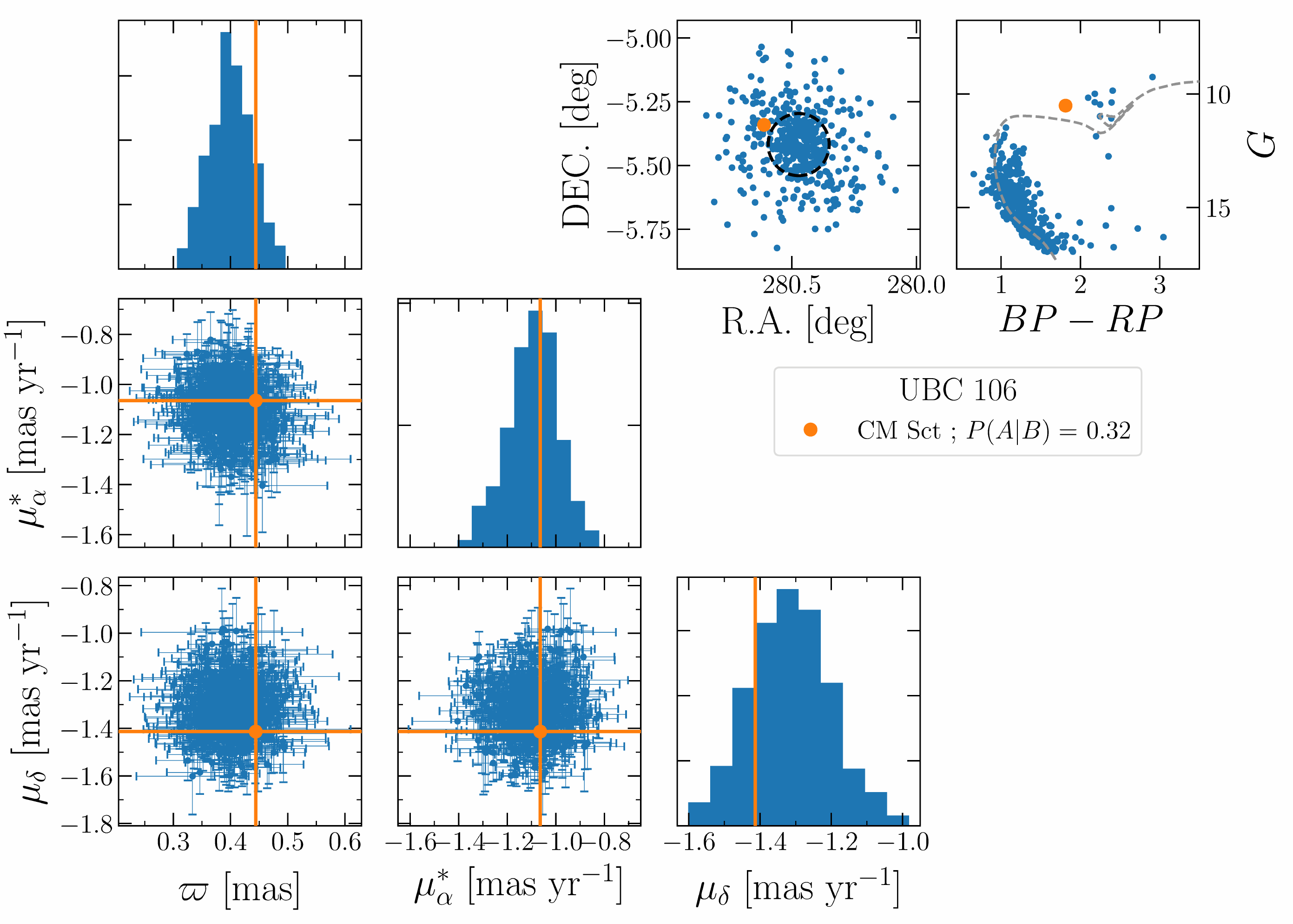}
\includegraphics[angle=0,scale=.29]{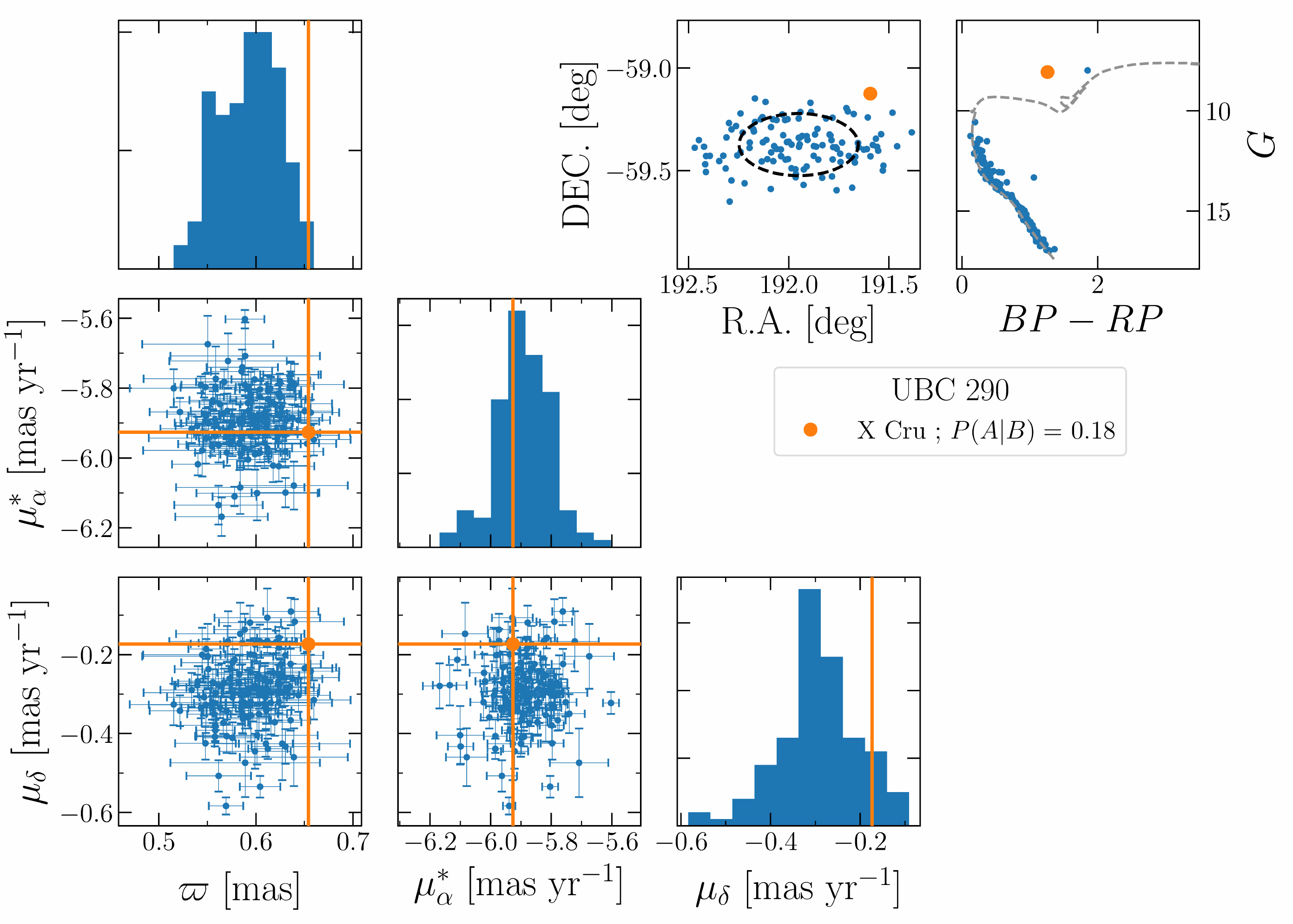}

\caption{
Distribution of the Gaia-based astrometry (parallaxes, proper motions, and positions) and colour-magnitude diagrams for the combos described in Section~\ref{sec:new}, for which a list of cluster members is provided by CG20. 
The information of the members (from CG20) is represented in blue, while the Cepheids properties are shown in orange. 
In the panels displaying the equatorial coordinates of the members, a black dashed line represent the clusters' $r_1$. 
PARSEC isochrones of solar metallicity are plotted in the colour-magnitude diagrams with grey dashed lines using the values derived by CG20 as a reference.}
\label{fig:multiplots}
\end{figure*}

\begin{figure*}
\includegraphics[angle=0,scale=.40]{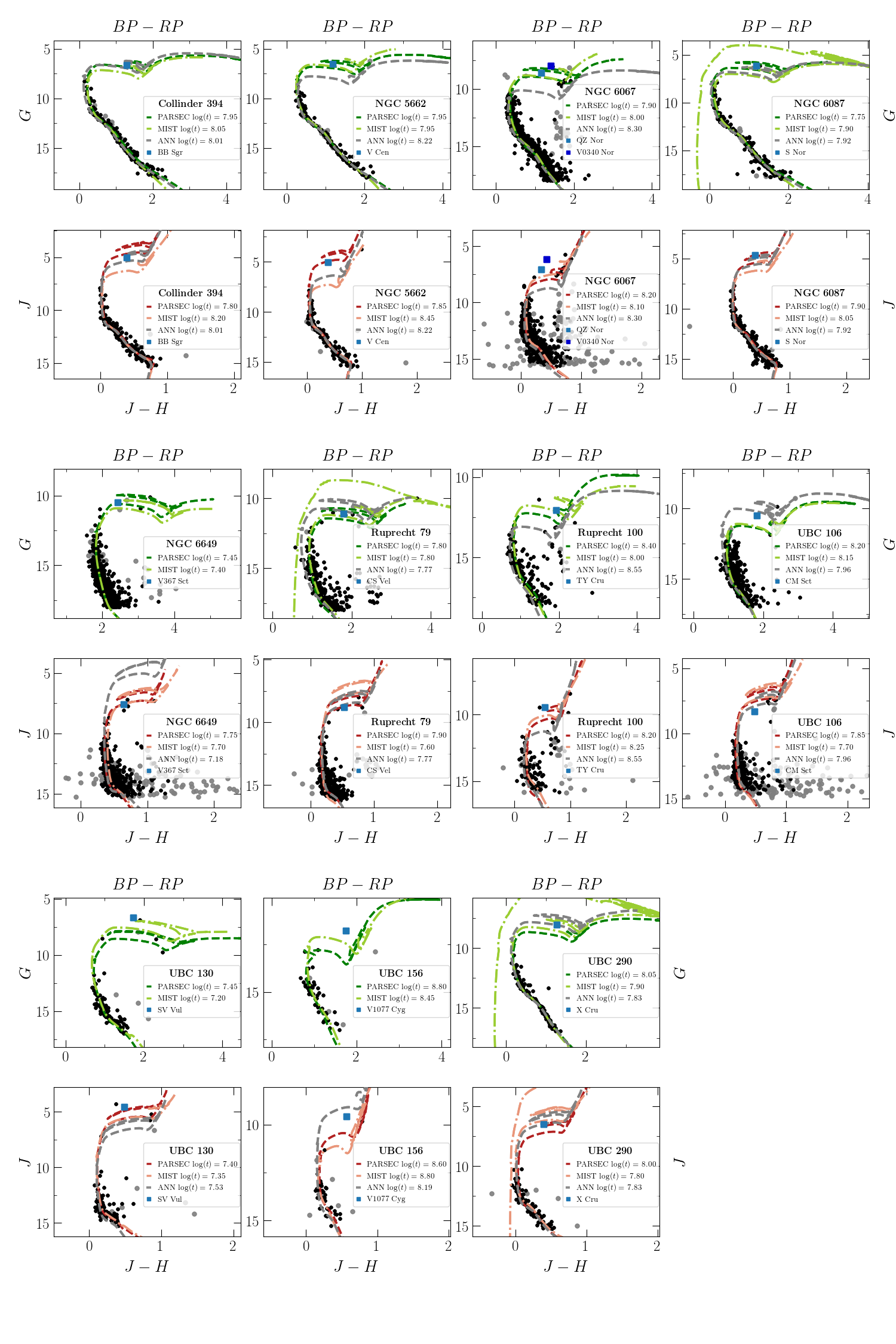}
\caption{Colour-magnitude diagrams of the clusters studied in Section~\ref{sec:ages}, with isochrones representing the results of Section~\ref{sec:methodology}. 
Cluster members from CG18b are shown with grey circles, whereas those used for the best-model determination are displayed in black. 
The outcomes of the ANN are plotted with PARSEC isochrones assuming solar metallicity (in grey), with the exception of the clusters marked as highly reddened in Table~\ref{tab:ages} (for the Gaia passband panels).  
}
\label{fig:allCMDs1}
\end{figure*}

\bsp	
\label{lastpage}
\end{document}